\documentclass[pdflatex,sn-mathphys-num]{sn-jnl}

\usepackage{graphicx}%
\usepackage{multirow}%
\usepackage{amsmath,amssymb,amsfonts}%
\usepackage{amsthm}%
\usepackage{mathrsfs}%
\usepackage[title]{appendix}%
\usepackage{textcomp}%
\usepackage{manyfoot}%
\usepackage{booktabs}%
\usepackage{algorithm}%
\usepackage{algorithmicx}%
\usepackage{algpseudocode}%
\usepackage{listings}%

\usepackage{url}
\usepackage[table,dvipsnames]{xcolor}
\definecolor{left}{HTML}{172869}     
\definecolor{left-center}{HTML}{0076BB}           
\definecolor{center}{HTML}{399043}         
\definecolor{right-center}{HTML}{AF6125}    
\definecolor{right}{HTML}{881C00}          
\definecolor{extreme-right}{HTML}{762A83} 

\usepackage{multirow}
\usepackage{adjustbox}
\usepackage{makecell}
\usepackage{booktabs}
\usepackage{longtable}
\usepackage{cleveref}

\usepackage{caption}
\usepackage{stfloats}
\usepackage{amsmath}

\usepackage{graphicx}
\usepackage{hyperref}

\begin{document}

\title[Platform Sorting Shapes Ideological Fragmentation in the Social Media Ecosystem]{Platform Sorting Drives Ideological Fragmentation in the Social Media Ecosystem}

\author*[1]{\fnm{Edoardo} \sur{Di Martino}}\email{edoardo.dimartino@uniroma1.it}
\author*[2]{\fnm{Alessandro} \sur{Galeazzi}}\email{alessandro.galeazzi@unipd.it}
\author*[3]{\fnm{Matteo} \sur{Cinelli}}\email{matteo.cinelli@uniroma1.it}
\author*[4]{\fnm{Michele} \sur{Starnini}}\email{michele.starnini@upf.edu}
\author*[3]{\fnm{Walter} \sur{Quattrociocchi}}\email{walter.quattrociocchi@uniroma1.it}

\affil*[1]{\orgdiv{Department of Social Sciences and Economics}, \orgname{Sapienza University of Rome}, \orgaddress{\street{P.le Aldo Moro, 5}, \postcode{00185}, \state{Rome}, \country{Italy}}}
\affil*[2]{\orgdiv{Department of Mathematics}, \orgname{University of Padova}, \orgaddress{\street{Via Trieste, 63}, \postcode{35121}, \state{Padua}, \country{Italy}}}
\affil*[3]{\orgdiv{Department of Computer Science}, \orgname{Sapienza University of Rome}, \orgaddress{\street{Viale Regina Elena, 295}, \postcode{00161}, \state{Rome}, \country{Italy}}}
\affil*[4]{\orgdiv{Department of Engineering}, \orgname{Universitat Pompeu Fabra}, \orgaddress{\postcode{08018}, \state{Barcelona}, \country{Spain}}}

\abstract{
Ideological asymmetries in online political communication are often studied as localized phenomena emerging within communities. 
Here, 
we show that 
fragmentation instead operates at the level of entire platforms, consistent with a process of platform sorting in which users increasingly align with ideologically congruent environments.
We analyze political information dynamics across Bluesky, Facebook, Reddit, Truth Social, Twitter/X, and YouTube during the 2020 and 2024 US presidential elections, combining measures of content sharing, engagement allocation, and user-level ideological orientation.
Across platforms, ideological fragmentation emerges consistently and persists over time. 
Platforms exhibit distinct ideological profiles that persist across the two election cycles, ranging from strongly left-leaning to strongly right-leaning environments. Longitudinal analyses further reveal limited ideological variability among persistent user cohorts, indicating that apparent changes within single platforms reflect ecosystem-level sorting rather than convergence toward neutrality.
Taken together, our results show that the dynamics of platform sorting is not a transient reaction to political events or moderation interventions, but a persistent structural feature of the social media ecosystem.

}
\maketitle
\section{Introduction}

The digital information ecosystem is increasingly characterized by uneven exposure to political content \cite{nyhan2023like, guess2023social, guess2023reshares}, contributing to growing audience fragmentation across online spaces \cite{gonzalez2023asymmetric, bakshy2015exposure}. 
Social media platforms were initially conceived as open infrastructures for connection and information exchange, but the shift toward individualized content streams, dynamics of engagement that prioritize interaction over accuracy or pluralism~\cite{sunstein2004democracy, voorveld2018engagement, sangiorgio2024followers, etta2023characterizing, zollo2026examining}, and the declining trust in traditional media institutions~\cite{gallup2018indicators, nic2018reuters}, has progressively redirected attention into ideologically aligned and fragmented contexts~\cite{bessi2015science, cinelli2021echo, terren2021echo, tucker2018social, hartmann2025systematic}. 
However, the scale at which this fragmentation operates remains debated, and whether it primarily reflects localized clustering within platforms or structural organization of the entire social media ecosystem is unclear.

Most research has interpreted fragmentation through the framework of echo chambers, understood as localized clusters of users in a platform repeatedly exposed to consonant viewpoints through algorithmic filtering and selective exposure~\cite{bakshy2015exposure, barbera2020social, cinelli2021echo}.  
Echo chambers have been associated with increased exposure to disinformation~\cite{diaz2023disinformation}, rising affective polarization~\cite{bail2018exposure, kubin2021role}, and the diffusion of antagonistic or extreme content~\cite{avalle2024persistent, tahmasbi2021go, league2020online}. 
While this perspective provides important insights into how individuals navigate political information online, it implicitly assumes that fragmentation remains confined to communities within ideologically heterogeneous platforms. However, the social media ecosystem has become increasingly fluid. New platforms continuously emerge, moderation and governance policies evolve~\cite{newell2016user,horta2021platform, ali2021understanding, monti2023online, schmitz2025volunteerism}, and users migrate across platforms in response to political, social, and technological changes~\cite{cava2023drivers, failla2024m, quelle2025bluesky}. 

Under these conditions, fragmentation may operate at a broader scale, reorganizing political attention across platforms rather than merely within them.  We define \emph{platform sorting} as the process through which users increasingly self-select into ideologically compatible platforms. Similar to the concept of partisan sorting in political science—the process by which the public aligns its party identification with its ideological positions~\cite{fiorina2004culture, levendusky2009partisan}— platform sorting describes the progressive alignment between users' political orientations and the platforms they frequent.
The result of the platform sorting dynamics are echo-platforms, characterized by coherent ideological majorities and asymmetric patterns of political information consumption~\cite{di2025ideological}. 
Therefore, a longitudinal perspective is crucial, as political events, moderation interventions, or migration waves might generate short-term fluctuations resembling structural changes. Whether the presence of echo platforms reflects transient rearrangements tied to specific political events or instead signals a persistent structural reconfiguration of the social media ecosystem remains an open empirical question.

In this work, we provide a longitudinal, multi-platform analysis of the persistence of echo platforms. 
We analyze political information dynamics across six social media platforms-- Bluesky, Facebook, Reddit, Truth Social, Twitter/X, and YouTube-- during the 2020 and 2024 US presidential elections, combining measures of content sharing, engagement allocation, and user-level ideological orientation. Within a unified analytical framework, our approach enables a direct comparison, moving beyond single-platform perspectives. Asymmetric distributions of political content, engagement, and user composition emerge consistently and remain stable across the two electoral periods. Platforms exhibit distinct ideological profiles, yet these differences reflect ecosystem-level fragmentation. Notably, these patterns do not characterize only platforms with a right-leaning user base: Bluesky, despite exhibiting a strongly left-leaning user base, displays structural patterns comparable to those observed elsewhere in the ecosystem.

Taken together, our results show that echo platforms are not episodic byproducts of polarization, such as temporary reactions to political shocks~\cite{kupferschmidt2024like, m2021political, bar2023new}, but stable configurations emerging from selective exposure~\cite{stroud2010polarization, spohr2017fake, brugnoli2019recursive, cardenal2019digital}. Viewed through the lens of platform sorting, polarization in online social media is better understood as a structural reorganization of ideological alignment across the ecosystem rather than a transient imbalance within platforms. This perspective reframes polarization as an emergent property of the contemporary social media ecosystem.

\section*{Results}
Our dataset comprises shared URLs from six social media platforms. 
We conducted a longitudinal analysis for Facebook, Reddit, YouTube, and Twitter (2020 and 2024 US presidential election cycles), while focusing exclusively on 2024 for more recent platforms (Bluesky and Truth Social).

We extract URLs from politically relevant posts, either through keyword-based collection, or, in the case of Reddit, by selecting communities primarily devoted to political discussion, where posts frequently consist solely of links to external news sources. 
We focus on URLs because they provide a common unit of analysis across otherwise heterogeneous environments: although platforms differ substantially in structure, affordances, and interaction mechanisms, the domains associated with shared URLs offer a consistent basis for cross-platform comparison. 
From the extracted domains, we only retain those labeled by Media Bias/Fact Check (MBFC; \url{https://mediabiasfactcheck.com}), a widely referenced \cite{pecile2025mapping, cinelli2021echo, stefanov2020predicting, flamino2023political} news rating agency, providing political bias and reliability labels for a large number of news outlet and websites. 
Restricting the analysis to labeled domains allows us to focus specifically on politically relevant information sources, while assigning a consistent set of ideological and reliability annotations across platforms. 

Table~\ref{tab:data_overview} in the Methods section reports the exact time frame and data volumes for each platform. 
While the observation windows differ slightly across platforms and election cycles, 
we obtain almost invariant results by using a unified observation window, see Figures~\ref{fig:common_windows},\ref{fig:shares_common_window},\ref{fig:random_eng_common_window} and Table~\ref{tab:delta_edgelists} in the Supplementary Information (SI).

\subsection*{\textbf{News sharing patterns}}

We begin by estimating the circulation of political information across platforms. As mentioned earlier, we analyze the political orientation and factual reliability of external news domains contained in user posts using standardized annotations from Media Bias/Fact Check: shared URLs are assigned to MBFC categories and, for each platform, their relative frequencies are computed.

\begin{figure}[tbp]
  \centering
  \includegraphics[width=.9\columnwidth]{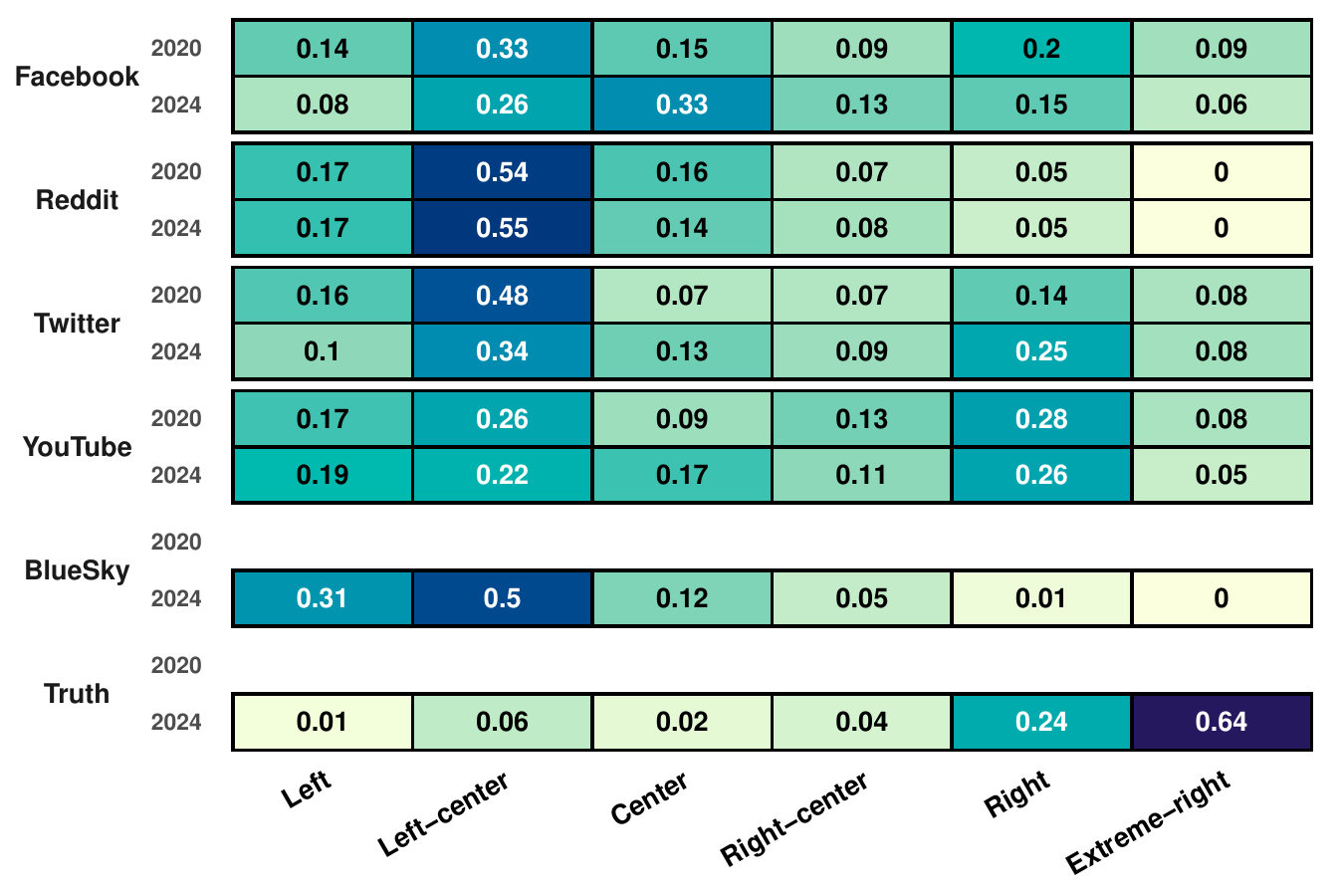}
\caption{\textbf{Platforms' news sharing patterns.}
Fraction of shared URLs linking to external news domains across platforms, stratified by political leaning according to MBFC labels. The figure highlights how content circulation differs systematically across platforms and evolves between the two election cycles, revealing distinct ideological compositions in the distribution of shared news sources. 
}
  \label{fig:weighted_diet}
\end{figure}

Figure~\ref{fig:weighted_diet} shows, for each platform, the fraction of shared URLs associated with news sources of different political leanings, in 2020 and 2024. 
For visual interpretability, domains classified as ``extreme-left" are collapsed into the ``left" category due to their small presence, with only 7 such domains identified between all platforms and time periods. 
We observe distinct ideological configurations between platforms: content circulation is not uniform, but reflects different news-sharing patterns across the ecosystem. 
Reddit shows a stable preference for left- and left-center domains across both election cycles. 
Facebook shows a shift toward more moderate content from 2020 to 2024, with a moderate decline in the volume of right-leaning and left-leaning sources circulating, and a relative doubling of center-leaning sources shared. This pattern appears to reflect the fact that center-leaning sources decrease at a slower rate than other categories on Facebook, rather than a substantial absolute increase in centrist content (see Table~\ref{tab:delta_leanings} and Figure~\ref{fig:diet_raw} in the SI).
YouTube displays a similar trend regarding the increase in center-leaning content shared (from $0.09$ to $0.17$), but the fraction of left- and right- leaning sources remains consistent across time. On the other hand, we observe a notable shift on Twitter, with a marked decrease of left and left-center domains between the two election cycles (from $0.16$ to $0.1$, and from $0.48$ to $0.34$, respectively), together with an 11\% increase in right-leaning content. 
Among the newer platforms, Bluesky displays a pronounced focus on left- and left-center domains, while 
Truth Social concentrates its news diet on right and, especially, extreme-right sources, with minimal attention to centrist or left-aligned content. 
With the possible exception of Facebook, these patterns do not seem to indicate convergence toward neutrality, reflecting instead a redistribution of ideological alignment across platforms.

\subsection*{\textbf{Preferential allocation of engagement}}

While platforms differ in the circulation of political content, the allocation of user engagement does not necessarily align with content circulation. To determine whether attention is distributed proportionally to content availability or reflects systematic preferences, we analyze observed engagement patterns against a platform-specific null model. The engagement received by a given source is defined as the sum of all interaction metrics available for each platform, as described in the Methods section. While these metrics differ across platforms due to structural differences and data-availability constraints, their distributions consistently exhibit similar heavy-tailed behavior (see Figure~\ref{fig:powerlaw_engagement} in the Supplementary Information), allowing meaningful comparisons when focusing on the fraction of total engagement received by different sources, rather than on raw engagement values.

Intuitively, our approach tests whether engagement is allocated proportionally to content supply. 
If users do not exhibit preferential behavior, sources belonging to a given political leaning should receive a share of total engagement proportional to their share of circulation. 
Formally, for each platform $p$, let $S_{p,\ell}$ denote the fraction of shared URLs linking to leaning $\ell$, and $E_{p,\ell}$ the fraction of total engagement they receive. In the absence of preferential allocation, engagement should mirror circulation,~$\mathbb{E}[E_{p,\ell}] = S_{p,\ell}$. We quantify deviations from this benchmark through the ratio~$R_{p,\ell} = E_{p,\ell}/S_{p,\ell}$, which is equal to one under proportional allocation, and departs from one when engagement is disproportionately allocated. 
To assess whether such deviations are statistically meaningful, we construct a null model in which political leaning labels are randomly reassigned across sources, while preserving all other empirical quantities, including the number of times each source is shared and the total engagement it receives. 
Repeating this procedure generates a reference distribution for $R_{p,\ell}$ under the hypothesis of no preferential allocation, against which we compare the observed values to identify over- and under-engaged. 

We then extend this framework to examine whether such deviations are driven by reliable or questionable sources: conditional on platform and political leaning, we compare the relative engagement of reliable and questionable sources and test whether any imbalance exceeds what would be expected under random assignment. 
By using a second null model in which reliability labels are randomly reshuffled within each platform and ideological group, we assess whether the observed differences in engagement between reliable and questionable sources reflect systematic preferences rather than random variation. 
See Methods for further details regarding both randomization processes.

\begin{figure}[tbp]
  \centering
  \includegraphics[width=\columnwidth]{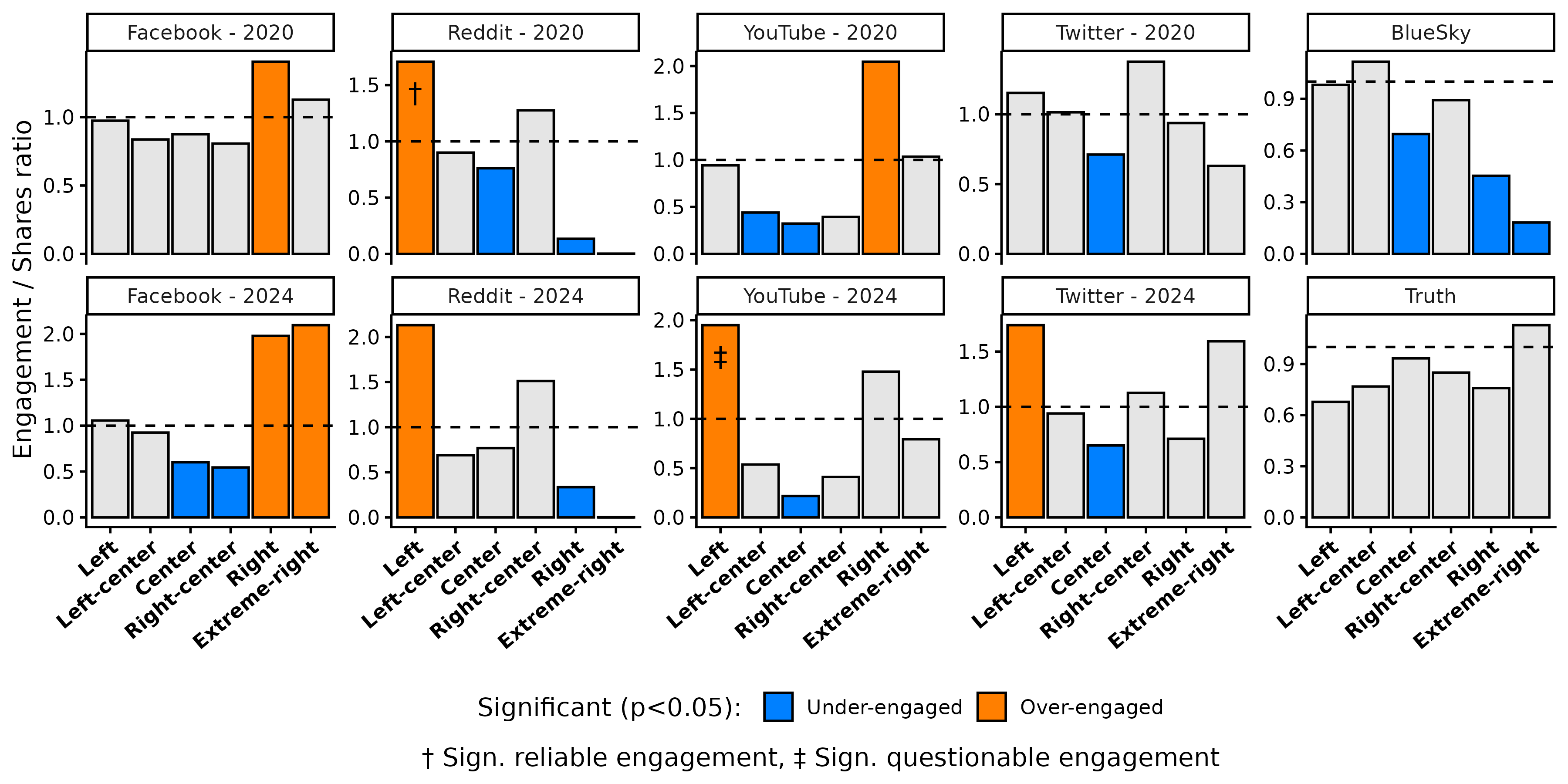}
\caption{\textbf{Engagement relative to content circulation.}
Engagement-to-share ratio $R_{p,\ell}$ across platforms and ideological leanings, together with the results of the permutation test. Values above (below) one indicate that engagement exceeds (falls below) expectations based on the circulation of news sources. Significant deviations from the null model highlight systematic over- or under-allocation of attention across ideological categories, revealing platform-specific patterns of preferential engagement. Single or double crosses indicate whether over-engagement is driven by reliable or questionable sources.}
  \label{fig:randomization_ratio}
\end{figure}

Figure \ref{fig:randomization_ratio} shows that engagement systematically departs from proportional allocations across all platforms, concentrating on specific political leanings that vary across on the platform and the time period considered. 
However, a common pattern across platforms (except for Truth), is the relative under-engagement of centrist sources, consistently receiving less attention than expected. 
On the contrary, partisan sources seem to be generally over-engaged with. 
Reddit shows a disproportionate and significant allocation of engagement toward left-leaning sources, while Facebook does so for right-leaning ones. YouTube presents a more polarized situation, with over-engagement of right-leaning sources in 2020 and left-leaning ones in 2024.
Interestingly, left-leaning over-engagement is driven by reliable sources on Reddit, while it is driven by questionable sources on YouTube. Twitter shows an engagement-to-share ratio $R_{p,\ell}$ above 1 for both left- and extreme-right sources in 2024, although only the over-engagement toward left-leaning sources is statistically significant. Truth Social exhibits an almost proportional allocation of engagement across ideological categories, likely due to the limited presence of non-right-leaning content on the platform (see Figure~\ref{fig:weighted_diet}).
Bluesky displays a complementary pattern: despite the relatively small circulation of right-leaning sources, these sources receive significantly less engagement than expected under the null model. Although their share of circulation appears close to zero in Figure~\ref{fig:weighted_diet} due to rounding, right-leaning sources are not entirely absent from the platform (21$,$179 URLs in total for Right-center, Right, and Extreme-right sources, see Figure~\ref{fig:diet_raw} in the SI).

This pattern reflects selective attention aligned with the platform’s prevailing ideological profile, consistent with platform-level sorting dynamics: certain platforms may not only attract a specific kind of audience, but also act as arenas in which content that aligns with the majority view can be amplified. 
Refer to Figure~\ref{fig:engagement_quest} in the SI for the complete results regarding the questionable vs reliable randomization.

\subsection*{\textbf{{Engagement similarity across platforms}}}

To examine whether different user bases converge or diverge in their engagement toward external sources, we compare the distribution of attention across platforms within a common domain space. For each platform, we identify the set of 20 domains receiving the highest engagement and consider the union of all such sets, resulting in 79 domains. 
Each platform is represented by a vector over the entries of this union set (i.e. joint domain space), with entries containing the fraction of total engagement directed toward each domain. 
The full list of the domains in the union set and the percentage of engagement for each source across platforms are reported in Table~\ref{tab:union_set} of the SI. 
Pairwise cosine similarity among vectors is used to quantify the alignment of engagement patterns across platforms. Supplementary Information Figures~\ref{fig:engagement_distribution},\ref{fig:cosine_robustness} provide additional robustness checks showing the consistency of results by using different thresholds.

\begin{figure}[tbp]
    \centering
    \includegraphics[width=\columnwidth]{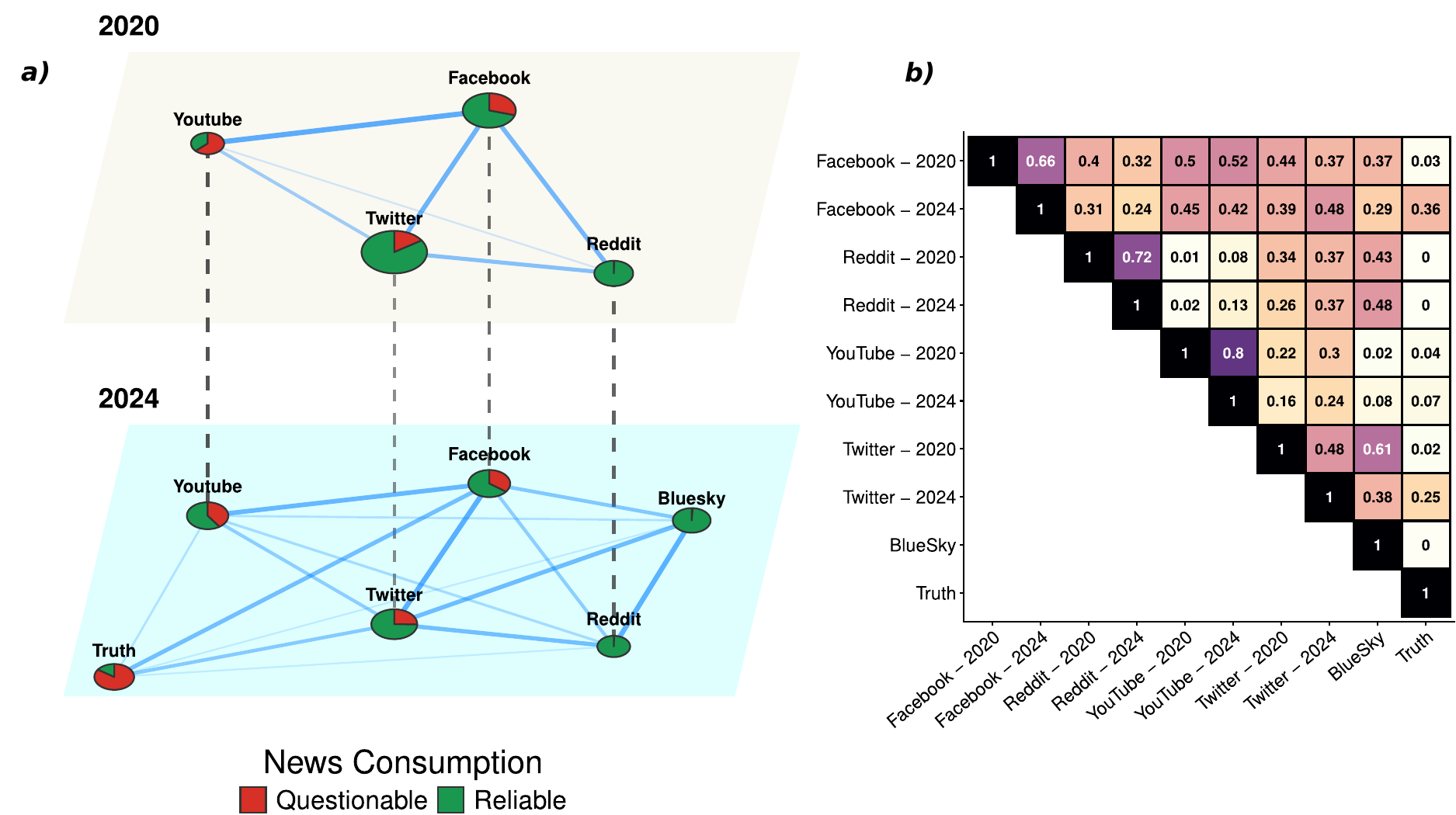}
\caption{\textbf{Engagement similarity across platforms and election cycles.}
Multiplex network representing cosine similarity between platforms based on their most engaged news domains (panel \textbf{\textit{a}}), with corresponding weighted adjacency matrix (panel \textbf{\textit{b}}). Node size is proportional to the total volume of shared links, while pie-chart colors indicate the fraction of engagement directed toward questionable versus reliable sources. The structure of the network highlights how engagement patterns cluster across platforms and evolve between 2020 and 2024, revealing the redistribution of attention across the social media ecosystem.}
    \label{fig:similarity_multiplex}
\end{figure}

We organize the resulting similarity (fully-connected) network into two layers, corresponding to 2020 and 2024, to examine both cross-platform cohesion and temporal persistence. 
The multiplex network in Figure~\ref{fig:similarity_multiplex} shows that, both in 2020 and 2024, Facebook and Twitter generally express a relative similarity with the other platforms, while Reddit and YouTube show a more specialized profile. By 2024, YouTube, Reddit, and Facebook express high self-similarity with their 2020 selves ($0.8$, $0.72$, and $0.66$, respectively), but, interestingly, Twitter's self-similarity is distinctly lower, with a value of $0.48$. Instead, the 2020-Twitter's most similar platform is Bluesky, which shows the highest similarity score among different platforms ($0.61$). Truth is much more similar to 2024-Twitter and 2024-Facebook than any other platforms.

Figure~\ref{fig:similarity_multiplex}(a) also illustrates the distribution of user engagement between questionable and reliable domains. While this proportion remains largely stable on Facebook and Reddit, other platforms exhibit distinct shifts. Twitter shows a slight uptick in questionable engagement, whereas YouTube recorded a notable decrease in 2024. This decline is primarily attributed to reduced engagement with Fox News, which previously commanded a staggering $49\%$ of YouTube's total engagement volume in 2020 (see Table \ref{tab:union_set}). Furthermore, Bluesky demonstrates negligible interaction with questionable content, contrasting sharply with Truth Social, where such domains capture a significantly larger share of attention. Collectively, these trends point toward an increasing specialization and personalization within the broader social media ecosystem, where engagement patterns are redistributed across platforms rather than within a single one.

\subsection*{\textbf{Ideological Profiles of Users}}

To characterize the ideological composition of the user bases across platforms, we estimate the political leaning of active users, defined as those who shared at least ten posts on a given platform within the considered election cycle. 
For each user, we compute an average leaning score based on the political orientation of the news domains they shared, using a continuous scale from $-1$ (extreme left) to $+1$ (extreme right) derived from MBFC annotations (see Methods for details). 

\begin{figure}[t]
  \centering
  \includegraphics[width=\columnwidth]{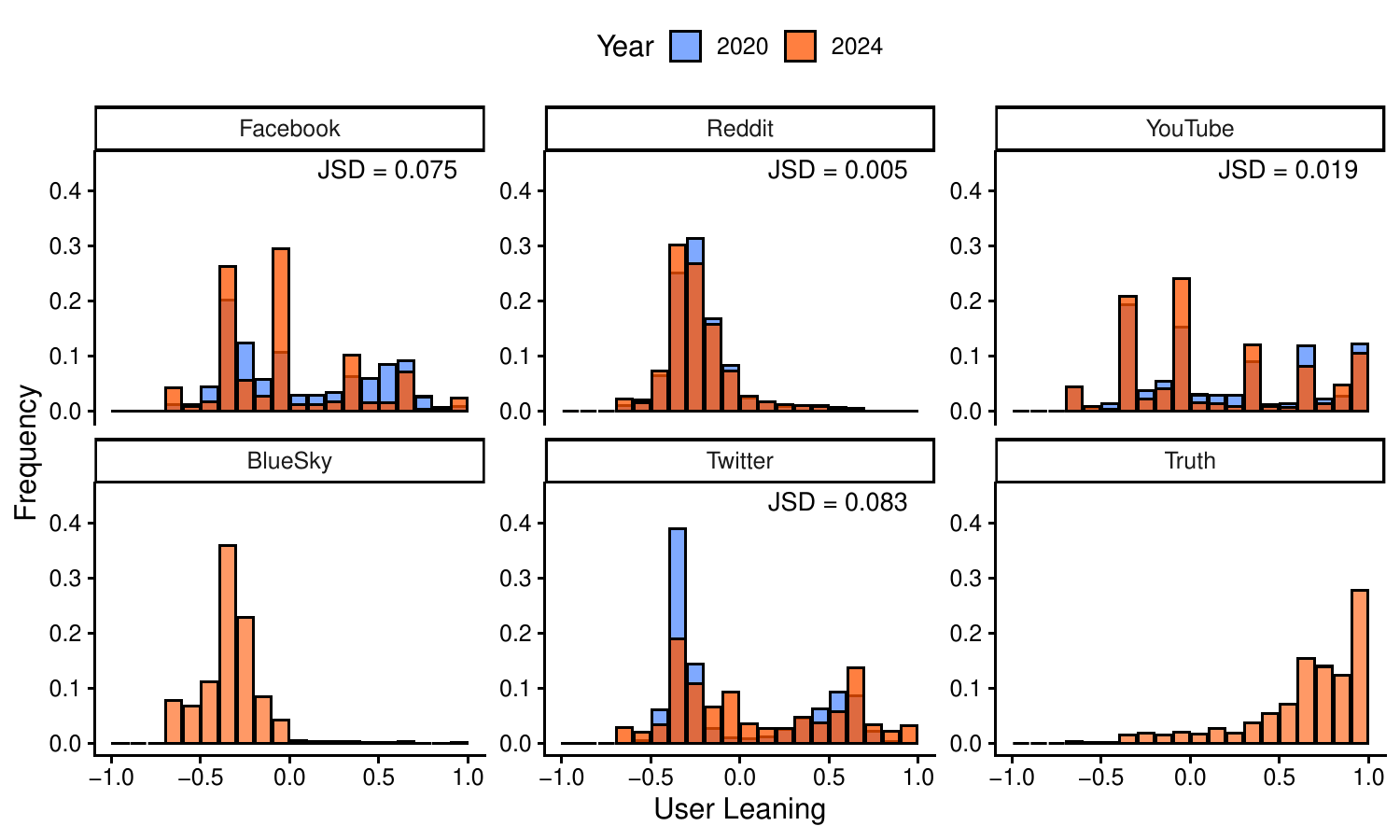}
\caption{\textbf{User ideological profiles across election cycles.}
Comparison of user ideological leaning distributions between 2020 and 2024 for platforms with longitudinal data. Each subplot shows the proportion of users across the ideological spectrum, ranging from $-1$ (extreme-left) to $+1$ (extreme-right). Jensen--Shannon divergence (JSD) values quantify the similarity between yearly distributions, with lower values indicating greater stability in the ideological composition of user bases.}
  \label{fig:leaning_comparison}
\end{figure}

Figure~\ref{fig:leaning_comparison} shows the distributions of user leaning for each platform separately. For platforms with longitudinal data (Facebook, Reddit, YouTube, and Twitter), the figure overlays the 2020 and 2024 distributions and reports the corresponding Jensen--Shannon divergence (JSD) between the two. For platforms without a 2020 baseline (Bluesky and Truth Social), only the 2024 distribution is shown. This representation enables direct comparison of ideological profiles across platforms and, where available, over time. Across the ecosystem, each platform exhibits a distinct ideological profile that remains (largely) stable over time. Reddit shows an almost invariant distribution between election cycles, with a clearly left-leaning user base, while YouTube displays modest changes and remains overall the more evenly distributed between the platforms. Facebook shows a moderate shift, expressing a similarly polarized situation in 2024 as it did in 2020, albeit with a bigger presence of neutral users. Twitter exhibits the most pronounced shift, with a reduction in left-leaning users and a corresponding increase in right-leaning ones, resulting in a more balanced bimodal configuration in 2024 compared to 2020. 
For platforms without longitudinal baselines, ideological asymmetries are also pronounced: Bluesky exhibits a strongly left-leaning user base, while Truth is heavily skewed toward the extreme right.

These findings show that, generally, the user leaning distributions mirror content exposure and engagement within platforms.
Notably, when the distributions of Twitter in 2024 and Bluesky are considered jointly, their combined shape resembles the ideological distribution observed on Twitter in 2020. This pattern is consistent with a redistribution of previously coexisting ideological groups across distinct platforms rather than with a reduction in ideological segregation itself.
While this comparison is qualitative, it suggests that ideological profiles previously coexisting on a single platform are now redistributed across multiple ones. 
This shift, combined with the fact that Twitter in 2020 and Bluesky in 2024 show nearly identical political engagement patterns, is consistent with the platform sorting dynamics of users.

\subsection*{\textbf{Ideological trajectories of persistent users}}

\begin{figure}[tbp]
  \centering
  \includegraphics[width=\columnwidth]{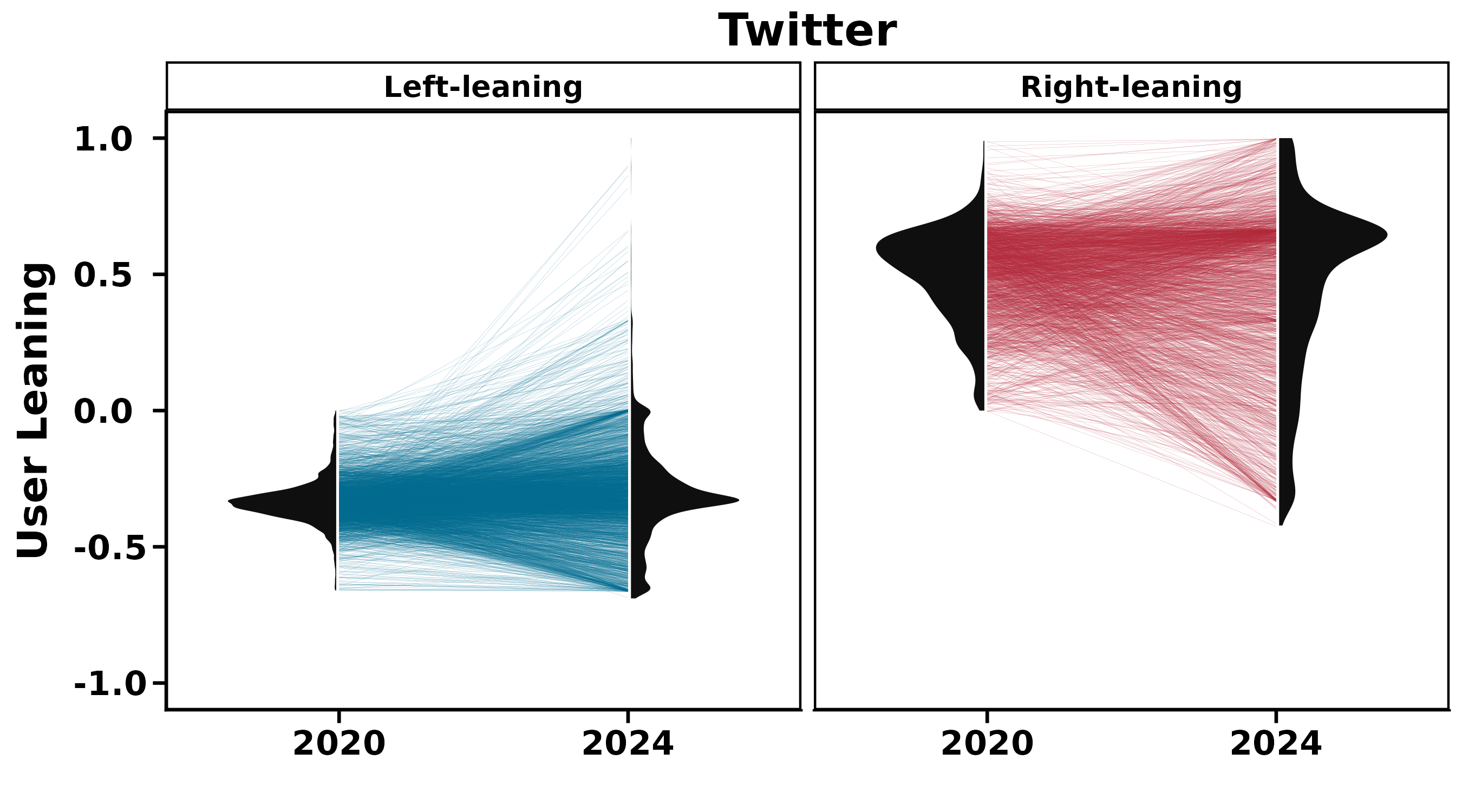}
  \caption{\textbf{Individual ideological dynamics on Twitter between 2020 and 2024.}
Empirical distribution of user-level ideological shifts. The figure highlights increasing dispersion within ideological groups without systematic drift toward the opposite pole, indicating stability in average ideological alignment despite growing heterogeneity.
}
  \label{fig:persistent_users}
\end{figure}

Our previous results point to an increasingly specialized social media ecosystem, raising the question of how the ideological orientation of users who remain active on the same platform evolves over time. 
Aggregate platform-level stability may arise through different mechanisms: it may reflect stable individual behavior, or result from user turnover, where departing users are replaced by new ones with similar orientations. To distinguish between these possibilities, we examine ideological trajectories at the individual level, focusing on persistent users—defined as individuals who posted at least ten times in both 2020 and 2024—on Twitter, the only platform providing a sufficiently large sample ($6{,}786$ users) for longitudinal analysis. Users are grouped into left- and right-leaning categories based on their 2020 leaning scores, using $0$ as a threshold. Figure~\ref{fig:persistent_users} shows their trajectories between 2020 and 2024, revealing dispersion within groups but no systematic drift toward the opposite pole, consistent with stable ideological alignment.

To assess whether observed changes exceed random fluctuation, we construct a null model capturing unstructured variability in individual leanings. 
For a user with average leaning $\theta_t$ and standard deviation $\sigma_t$ (both defined over the set of tweets at time $t$), 
we consider $\theta_{t+1} = \theta_t + \epsilon$, 
where $\epsilon \sim \mathcal{N}(0, \sigma_t)$. 
Simulated values are clipped to $[-1, +1]$. This procedure preserves the empirical distribution of initial leanings while removing structured or directional change, providing a baseline for comparison.
Empirical trajectories differ systematically from the null model. Transition probabilities (Table~\ref{tab:leaning_change}) show that most users remain in the same ideological category, with persistence rates of $0.962$ for left-leaning and $0.874$ for right-leaning users. The null model predicts lower stability for left-leaning users and comparable (albeit slightly higher) stability for right-leaning ones, suggesting that that individual ideological trajectories are not driven by random fluctuations, but reflects actual individual preferences.

\begin{table}[tbp]
\centering
\caption{
\textbf{Transition probabilities between ideological categories for persistent users.}
Transition probabilities from ideological group membership in 2020 to 2024 for empirical data and for the null model. Null model values are averaged over $10^3$ realizations. 
}
\begin{tabular}{lcc|cc}
\toprule
& \multicolumn{2}{c}{\textbf{Empirical transitions}} & \multicolumn{2}{c}{\textbf{Null model transitions}} \\
\cmidrule(lr){2-3} \cmidrule(lr){4-5}
\textbf{2020 group} & Left-leaning & Right-leaning & Left-leaning & Right-leaning \\
\midrule
Left-leaning  & 0.962  & 0.038 &  0.909 & 0.091 \\
Right-leaning & 0.126 & 0.874 & 0.119 & 0.881 \\
\bottomrule
\end{tabular}
\label{tab:leaning_change}
\end{table}

\section*{Conclusions}

Building on the concept of echo platforms, we showed that ideological asymmetries in exposure, engagement, and user composition remain stable over time, rather than converging toward neutrality. 
Longitudinal analyses suggest that this stability does not emerge from random turnover alone, as users active across both election cycles display very limited ideological drift, showing substantial persistence within ideological groups. 
Taken together, our results support the presence of a platform sorting dynamics across the social media ecosystem. Rather than disappearing, ideological alignments and engagement patterns reorganize across platforms, producing an increasingly fragmented ecosystem characterized by coherent ideological majorities and selective amplification of aligned content. 
With this perspective, polarization is better understood not simply as localized fragmentation within individual platforms, but as an ecosystem-level redistribution of ideological alignment.

At the same time, our findings should be interpreted in light of their limitations. While focusing on political discourse during the US presidential elections allows us to observe user behavior in a context where partisan and political dynamics are especially prominent, the analysis is necessarily tied to a specific topic and time frame. 
As such, the patterns observed here may not generalize to other domains or less polarized contexts. 
In addition, some spaces and dynamics for political debate may remain unobservable. Indeed, niche platforms frequently emerge, decline, or reorganize, and several others that played relevant roles in previous phases of online political fragmentation (e.g., Voat, Parler, Gab~\cite{di2025ideological, di2024users, m2021political, dehghan2022politicization}) have been either shut down or have become inaccessible during the course of this research, making longitudinal comparisons unfeasible. 
Finally, the empirical data available do not allow strong causal claims regarding the mechanisms driving platform sorting, user migration, or selective exposure, but only support the identification of robust, large-scale regularities, consistent with ecosystem-level fragmentation dynamics.

These findings have implications for both measurement and intervention. As a consequence of platform sorting dynamics, local reductions in polarization within individual platforms may coexist with, or even conceal, broader fragmentation within an increasingly specialized digital ecosystem. As such, strategies designed to reduce polarization within individual platforms may have limited impact if users can maintain stable preferences while reorganizing across platforms. 
Understanding political polarization, therefore, requires an ecosystem-level perspective that accounts for the interaction between user behavior, platform structure, and the distribution of attention across the broader digital environment. We further argue that, as recommendation and generative artificial intelligence become increasingly integrated into information access and discovery, they may reinforce fragmentation by continuously adapting content exposure to existing ideological positions~\cite{piao2023human, schroeder2026malicious}.

\clearpage

\section*{Methods}
\subsection*{\textbf{Data and engagement metrics}}

\begin{table}[htbp]
\centering
\begin{tabular*}{1.1\textwidth}{@{\extracolsep{\fill}} llrrrrrrcc}
\toprule
Platform & Year & $n_u$ & $n_d$ & $q_u$ & $q_e$ & $N$ & $N_{\text{active}}$ & From & To \\
\midrule

\multirow{2}{*}{Facebook}
  & 2020 & 5,167,018 & 4,084 & 0.239 & 0.295 & 150,727 & 39,290 & 25/05/2020 & 16/11/2020 \\
  & 2024 & 662,520 & 3,018 & 0.167 & 0.353 & 16,084 & 4,257 & 01/05/2024 & 15/11/2024 \\
\addlinespace

\multirow{2}{*}{Reddit}
  & 2020 & 372,196 & 2,571 & 0.044 & 0.005 & 39,005 & 3,030 & 01/01/2020 & 31/12/2020 \\
  & 2024 & 112,844 & 1,777 & 0.036 & 0.004 & 23,985 & 1,589 & 01/01/2024 & 31/12/2024 \\
\addlinespace

\multirow{2}{*}{Twitter}
  & 2020 & 69,579,321 & 3,555 & 0.183 & 0.158 & 3,629,917 & 653,349 & 01/06/2020 & 03/11/2020 \\
  & 2024 & 1,450,124 & 3,642 & 0.304 & 0.253 & 219,997 & 23,431 & 01/01/2024 & 06/11/2024 \\
\addlinespace

\multirow{2}{*}{YouTube}
  & 2020 & 117,419 & 1,267 & 0.348 & 0.640 & 2,500 & 662 & 01/06/2020 & 16/12/2020 \\
  & 2024 & 583,011 & 1,416 & 0.253 & 0.403 & 1,642 & 1,017 & 01/06/2020 & 31/12/2020 \\
\addlinespace

\multirow{2}{*}{Bluesky}
  & 2020 & -- & -- & -- & -- & -- & -- & -- \\
  & 2024 & 305,448 & 2,227 & 0.029 & 0.009 & 55,541 & 4,353 & 02/01/2024 & 31/12/2024 \\
\addlinespace

\multirow{2}{*}{Truth}
  & 2020 & -- & -- & -- & -- & -- & -- & -- \\
  & 2024 & 171,748 & 1,521 & 0.798 & 0.852 & 7,102 & 1,192 & 01/01/2024 & 29/10/2024 \\
\addlinespace

\bottomrule
\end{tabular*}
\caption{Summary of data by platform and year. Columns report: number of labeled URL links ($n_u$), number of unique labeled domains ($n_d$), fraction of questionable sources shared ($q_u$) and fraction of engagement received by questionable sources ($q_e$), number of users sharing at least one labeled URL ($N$), number of users sharing at least five labeled URLs ($N_{\text{active}}$), and the dates of the first and last posts available (From, To).}
\label{tab:data_overview}
\end{table}

In this section, we describe the procedures for data collection and preprocessing, together with an overview of the datasets used in our analysis, and a description of the engagement metrics employed in the analysis of each platform. All datasets were filtered to retrieve politically relevant content related to the 2020 and 2024 US presidential elections. Depending on data availability and allowed collection methods, this was achieved either through keyword-based retrieval or by selecting politically focused communities when keyword filtering was not feasible. The set of keywords was largely consistent across platforms and included candidate names and election-related terms, as detailed in the following lines. Table \ref{tab:data_overview} contains further details regarding the metrics and time frames present for each data set.

\textbf{Bluesky:} Bluesky is a decentralized social media platform built on the AT Protocol. The dataset was obtained by collecting every posts published between January and December 2024 containing at least one of the following keywords: \texttt{\{Trump | Biden | Kamala | Harris | US Elections\}}, from which relevant URLs were extracted. The data collection procedure was performed during the month of April 2025. 
For engagement metrics on Bluesky, we consider the number of likes, replies, quotes and reposts associated to a post.

\textbf{Reddit:} We extracted relevant URLs from every post published in the subreddit \textit{r/Politics} between January and December 2020 and between January and December 2024, collected through the Pushshift archive~\cite{baumgartner2020pushshift}. Subreddits are user-created communities organized around specific topics, where users can submit posts, comment, and vote on content. For each Reddit post, we consider its score -defined as the number of upvotes (akin to likes) on a post, minus its number of downvotes (akin to dislikes)- as engagement metric, together with the number of comments under the post.

\textbf{Twitter/X:} The 2020 dataset is based on the collection described by Flamino \emph{et al.}~\cite{flamino2023political}, obtained via the Twitter Search API using candidate names (\texttt{\{Trump | Biden\}}) as keywords, where the relevant URLs were extracted from a collection of 126 million tweets posted between June and November 2020. For 2024, we collect URLs by leveraging the dataset introduced by Balasubramanian \emph{et al.}~\cite{balasubramanian2024public}, which comprises approximately 22 million posts related to the US presidential election between May and July 2024. Due to differences in data availability, for the 2020 dataset, we consider the number of retweets and favourites that a tweet receives as engagement metrics, while for the 2024 dataset, we utilize retweets, favourites, and quotes pertaining to a tweet.

\textbf{YouTube:} We analyze URLs extracted from the descriptions of approximately 270$,$000 videos collected through the YouTube Data API between June and December 2020. Videos were identified using the same keyword-based queries applied to other platforms, and additional related videos were collected through the recommendation network. Among related videos, only videos containing  \texttt{\{Trump\}} or \texttt{\{Biden\}} in the title or description were retained. 
Given YouTube API changes that do not allow the crawling of related videos as a way to expand the results of keywords search, at the time of data collection for the 2024 data (August 2025), instead of performing a keyword search, we first collect every video uploaded between June and December 2024 by the same channels considered in 2020. Using this approach, we were able to recover videos from 7$,$002 of the 12$,$991 original channels. To further extend coverage, we additionally collected videos from 1$,$480 news outlets that were not part of the original channel list, identified by matching Media Bias/Fact Check (MBFC) news outlets to their corresponding YouTube channels. All videos were subsequently filtered for content related to the 2024 Elections, using the usual keywords \texttt{\{Trump | Biden | Kamala | Harris | US Elections\}}. For both datasets, we employ the number of likes and the number of comments under a video as engagement metrics.

\textbf{Truth Social:} Truth Social is a social media platform launched in 2022 and widely used for political communication. We use a dataset capturing activity related to the 2024 US presidential election, comprising approximately 1.5 million posts published between February and October 2024, originally collected by Shah \emph{et al.} \cite{shah2024unfiltered}. Engagement metrics considered for Truth were the number of retruths (akin to retweets), quotes, and likes pertaining to a post.

\textbf{Facebook:} Data related to the 2020 Elections were collected from CrowdTangle, a Meta owned tool for data collection that was active until August 14 2024. CrowdTangle provided access to a limited set of accounts. For Facebook, the dataset included over 7 million public Facebook pages and verified profiles. Specifically, it comprised all public Facebook pages with more than 50,000 likes, all public groups with over 95,000 members, all US-based public groups with more than 2,000 members, and all verified profiles. The URLs analyzed in the study were extracted from 21 million Facebook posts collected using the CrowdTangle service between June and December 2020, which were identified based on searches using the presidential candidates names \texttt{\{Trump | Biden\}} as keywords, spanning the period from May to November 2020. Following the discontinuation of CrowdTangle, data for the 2024 election were collected using the \textit{Meta Content Library}, a web-based research tool designed as its replacement. Using this tool, in December 2025, we collected every available post with an associated URL based on the same keyword strategy adopted for Bluesky: \texttt{\{Trump | Biden | Kamala | Harris | US Elections\}}. For both the 2020 and 2024 Facebook data, we employ the number of comments, shares, and reactions under a post as engagement metrics.

\subsection*{\textbf{Labeling of news sources}}

News domains were labeled using annotations from Media Bias/Fact Check (MBFC; \url{https://mediabiasfactcheck.com}), an external source that provides assessments of political bias and factual reporting for a large set of news outlets. The list of labeled news outlets and domains adopted in this study was collected in October 2024 and includes categorical political bias scores ranging from extreme-left to extreme-right. In addition to political bias categories, MBFC assigns a ``questionable'' label to outlets that, according to their coding scheme, exhibit characteristics such as extreme bias, promotion of conspiratorial narratives, limited sourcing practices, or low transparency. We use these annotations exclusively as standardized metadata to enable comparative analysis across platforms, without independently evaluating the editorial practices of individual outlets.

\subsection*{\textbf{Null models for engagement allocation}}
To assess whether user engagement is distributed proportionally to content circulation, we compare the observed fraction of engagement received by different categories of sources with their relative presence in each platform's news diet. For a platform $p$ and political leaning $\ell$, let $S_{p,\ell}$ denote the fraction of shared URLs linking to sources with leaning $\ell$, and let $E_{p,\ell}$ denote the fraction of total engagement received by those sources. Under proportional allocation of attention, engagement is expected to mirror content circulation, i.e., $E_{p,\ell} = S_{p,\ell}$. We quantify deviations from this benchmark through the engagement-to-share ratio

\[
R_{p,\ell} = \frac{E_{p,\ell}}{S_{p,\ell}},
\]

where values above one indicate over-engagement relative to circulation, and values below one indicate under-engagement.

To evaluate statistical significance, we construct a permutation-based null model in which political leaning labels are randomly reassigned across sources while preserving all other empirical quantities, including the number of URLs shared for each source and the total engagement received. For each randomization, $R_{p,\ell}$ is recomputed, generating a null distribution corresponding to the absence of preferential engagement allocation across ideological categories. The procedure is repeated $10^3$ times independently for each platform. Empirical values are then compared against the corresponding null distributions to identify statistically significant cases of over- and under-engagement.

We further examine whether deviations from proportional allocation are preferentially associated with reliable or questionable sources. Conditional on platform $p$ and political leaning $\ell$, we compute the engagement-to-share ratios separately for reliable ($R_r$) and questionable ($R_q$) sources, and define their log-difference as

\[
\Delta = \log(R_r) - \log(R_q).
\]

Positive values of $\Delta$ indicate relatively greater engagement toward reliable sources, whereas negative values indicate greater engagement toward questionable sources. Statistical significance is assessed through a second permutation-based null model in which reliability labels are randomly reassigned within each platform--leaning group while preserving both total engagement and source circulation. For each realization, $\Delta$ is recomputed, generating an empirical null distribution against which the observed value is compared. The full results of this second permutation test can be referred to in Figure~\ref{fig:engagement_quest} of the SI.

\subsection*{\textbf{Inferring accounts' ideological leaning}}

To estimate the ideological leaning of individual accounts, we assign a numerical score to each external news domain based on its MBFC political bias label. Scores range from $-1$ (extreme-left) to $+1$ (extreme-right), with intermediate categories mapped as follows: $-0.66$ (left), $-0.33$ (left-center), $0$ (center/least biased), $0.33$ (right-center), and $0.66$ (right).

For an account $i$ that shared $n$ URLs linking to external domains, denoted by $C_i = \{c_1, c_2, \dots, c_n\}$, each shared domain $c_j$ is associated with its corresponding bias score. The ideological leaning of account $i$ is defined as the average bias score across all shared domains:

\begin{equation*}
x_i = \frac{1}{n}\sum_{j=1}^{n} c_j .
\end{equation*}

The resulting value lies in the interval $[-1,1]$, where negative (positive) values indicate left-leaning (right-leaning) sharing behavior. This operationalization is consistent with prior work showing that patterns of selective exposure to news sources provide a reliable proxy for political orientation in online environments~\cite{stroud2010polarization, cinelli2021echo, flamino2023political}.

\subsection*{\textbf{Acknowledgments}}

We gratefully thank Matteo Flamino for providing access to the engagement metrics associated with the 2020 Twitter dataset.
M.S. acknowledges support from Grants No.
RYC2022-037932-I and CNS2023-144156 funded by
MCIN/AEI/10.13039/501100011033 and the European
Union NextGenerationEU/PRTR.


\begin{thebibliography}{52}
\ifx \bisbn   \undefined \def \bisbn  #1{ISBN #1}\fi
\ifx \binits  \undefined \def \binits#1{#1}\fi
\ifx \bauthor  \undefined \def \bauthor#1{#1}\fi
\ifx \batitle  \undefined \def \batitle#1{#1}\fi
\ifx \bjtitle  \undefined \def \bjtitle#1{#1}\fi
\ifx \bvolume  \undefined \def \bvolume#1{\textbf{#1}}\fi
\ifx \byear  \undefined \def \byear#1{#1}\fi
\ifx \bissue  \undefined \def \bissue#1{#1}\fi
\ifx \bfpage  \undefined \def \bfpage#1{#1}\fi
\ifx \blpage  \undefined \def \blpage #1{#1}\fi
\ifx \burl  \undefined \def \burl#1{\textsf{#1}}\fi
\ifx \doiurl  \undefined \def \doiurl#1{\url{https://doi.org/#1}}\fi
\ifx \betal  \undefined \def \betal{\textit{et al.}}\fi
\ifx \binstitute  \undefined \def \binstitute#1{#1}\fi
\ifx \binstitutionaled  \undefined \def \binstitutionaled#1{#1}\fi
\ifx \bctitle  \undefined \def \bctitle#1{#1}\fi
\ifx \beditor  \undefined \def \beditor#1{#1}\fi
\ifx \bpublisher  \undefined \def \bpublisher#1{#1}\fi
\ifx \bbtitle  \undefined \def \bbtitle#1{#1}\fi
\ifx \bedition  \undefined \def \bedition#1{#1}\fi
\ifx \bseriesno  \undefined \def \bseriesno#1{#1}\fi
\ifx \blocation  \undefined \def \blocation#1{#1}\fi
\ifx \bsertitle  \undefined \def \bsertitle#1{#1}\fi
\ifx \bsnm \undefined \def \bsnm#1{#1}\fi
\ifx \bsuffix \undefined \def \bsuffix#1{#1}\fi
\ifx \bparticle \undefined \def \bparticle#1{#1}\fi
\ifx \barticle \undefined \def \barticle#1{#1}\fi
\bibcommenthead
\ifx \bconfdate \undefined \def \bconfdate #1{#1}\fi
\ifx \botherref \undefined \def \botherref #1{#1}\fi
\ifx \url \undefined \def \url#1{\textsf{#1}}\fi
\ifx \bchapter \undefined \def \bchapter#1{#1}\fi
\ifx \bbook \undefined \def \bbook#1{#1}\fi
\ifx \bcomment \undefined \def \bcomment#1{#1}\fi
\ifx \oauthor \undefined \def \oauthor#1{#1}\fi
\ifx \citeauthoryear \undefined \def \citeauthoryear#1{#1}\fi
\ifx \endbibitem  \undefined \def \endbibitem {}\fi
\ifx \bconflocation  \undefined \def \bconflocation#1{#1}\fi
\ifx \arxivurl  \undefined \def \arxivurl#1{\textsf{#1}}\fi
\csname PreBibitemsHook\endcsname

\bibitem[\protect\citeauthoryear{Nyhan et~al.}{2023}]{nyhan2023like}
\begin{barticle}
\bauthor{\bsnm{Nyhan}, \binits{B.}},
\bauthor{\bsnm{Settle}, \binits{J.}},
\bauthor{\bsnm{Thorson}, \binits{E.}},
\bauthor{\bsnm{Wojcieszak}, \binits{M.}},
\bauthor{\bsnm{Barber{\'a}}, \binits{P.}},
\bauthor{\bsnm{Chen}, \binits{A.Y.}},
\bauthor{\bsnm{Allcott}, \binits{H.}},
\bauthor{\bsnm{Brown}, \binits{T.}},
\bauthor{\bsnm{Crespo-Tenorio}, \binits{A.}},
\bauthor{\bsnm{Dimmery}, \binits{D.}}, \betal:
\batitle{Like-minded sources on facebook are prevalent but not polarizing}.
\bjtitle{Nature}
\bvolume{620}(\bissue{7972}),
\bfpage{137}--\blpage{144}
(\byear{2023})
\end{barticle}
\endbibitem

\bibitem[\protect\citeauthoryear{Guess et~al.}{2023a}]{guess2023social}
\begin{barticle}
\bauthor{\bsnm{Guess}, \binits{A.M.}},
\bauthor{\bsnm{Malhotra}, \binits{N.}},
\bauthor{\bsnm{Pan}, \binits{J.}},
\bauthor{\bsnm{Barber{\'a}}, \binits{P.}},
\bauthor{\bsnm{Allcott}, \binits{H.}},
\bauthor{\bsnm{Brown}, \binits{T.}},
\bauthor{\bsnm{Crespo-Tenorio}, \binits{A.}},
\bauthor{\bsnm{Dimmery}, \binits{D.}},
\bauthor{\bsnm{Freelon}, \binits{D.}},
\bauthor{\bsnm{Gentzkow}, \binits{M.}}, \betal:
\batitle{How do social media feed algorithms affect attitudes and behavior in an election campaign?}
\bjtitle{Science}
\bvolume{381}(\bissue{6656}),
\bfpage{398}--\blpage{404}
(\byear{2023})
\end{barticle}
\endbibitem

\bibitem[\protect\citeauthoryear{Guess et~al.}{2023b}]{guess2023reshares}
\begin{barticle}
\bauthor{\bsnm{Guess}, \binits{A.M.}},
\bauthor{\bsnm{Malhotra}, \binits{N.}},
\bauthor{\bsnm{Pan}, \binits{J.}},
\bauthor{\bsnm{Barber{\'a}}, \binits{P.}},
\bauthor{\bsnm{Allcott}, \binits{H.}},
\bauthor{\bsnm{Brown}, \binits{T.}},
\bauthor{\bsnm{Crespo-Tenorio}, \binits{A.}},
\bauthor{\bsnm{Dimmery}, \binits{D.}},
\bauthor{\bsnm{Freelon}, \binits{D.}},
\bauthor{\bsnm{Gentzkow}, \binits{M.}}, \betal:
\batitle{Reshares on social media amplify political news but do not detectably affect beliefs or opinions}.
\bjtitle{Science}
\bvolume{381}(\bissue{6656}),
\bfpage{404}--\blpage{408}
(\byear{2023})
\end{barticle}
\endbibitem

\bibitem[\protect\citeauthoryear{Gonz{\'a}lez-Bail{\'o}n et~al.}{2023}]{gonzalez2023asymmetric}
\begin{barticle}
\bauthor{\bsnm{Gonz{\'a}lez-Bail{\'o}n}, \binits{S.}},
\bauthor{\bsnm{Lazer}, \binits{D.}},
\bauthor{\bsnm{Barber{\'a}}, \binits{P.}},
\bauthor{\bsnm{Zhang}, \binits{M.}},
\bauthor{\bsnm{Allcott}, \binits{H.}},
\bauthor{\bsnm{Brown}, \binits{T.}},
\bauthor{\bsnm{Crespo-Tenorio}, \binits{A.}},
\bauthor{\bsnm{Freelon}, \binits{D.}},
\bauthor{\bsnm{Gentzkow}, \binits{M.}},
\bauthor{\bsnm{Guess}, \binits{A.M.}}, \betal:
\batitle{Asymmetric ideological segregation in exposure to political news on facebook}.
\bjtitle{Science}
\bvolume{381}(\bissue{6656}),
\bfpage{392}--\blpage{398}
(\byear{2023})
\end{barticle}
\endbibitem

\bibitem[\protect\citeauthoryear{Bakshy et~al.}{2015}]{bakshy2015exposure}
\begin{barticle}
\bauthor{\bsnm{Bakshy}, \binits{E.}},
\bauthor{\bsnm{Messing}, \binits{S.}},
\bauthor{\bsnm{Adamic}, \binits{L.A.}}:
\batitle{Exposure to ideologically diverse news and opinion on facebook}.
\bjtitle{Science}
\bvolume{348}(\bissue{6239}),
\bfpage{1130}--\blpage{1132}
(\byear{2015})
\end{barticle}
\endbibitem

\bibitem[\protect\citeauthoryear{Sunstein}{2004}]{sunstein2004democracy}
\begin{barticle}
\bauthor{\bsnm{Sunstein}, \binits{C.R.}}:
\batitle{Democracy and filtering}.
\bjtitle{Communications of the ACM}
\bvolume{47}(\bissue{12}),
\bfpage{57}--\blpage{59}
(\byear{2004})
\end{barticle}
\endbibitem

\bibitem[\protect\citeauthoryear{Voorveld et~al.}{2018}]{voorveld2018engagement}
\begin{barticle}
\bauthor{\bsnm{Voorveld}, \binits{H.A.}},
\bauthor{\bsnm{Van~Noort}, \binits{G.}},
\bauthor{\bsnm{Muntinga}, \binits{D.G.}},
\bauthor{\bsnm{Bronner}, \binits{F.}}:
\batitle{Engagement with social media and social media advertising: The differentiating role of platform type}.
\bjtitle{Journal of advertising}
\bvolume{47}(\bissue{1}),
\bfpage{38}--\blpage{54}
(\byear{2018})
\end{barticle}
\endbibitem

\bibitem[\protect\citeauthoryear{Sangiorgio et~al.}{2024}]{sangiorgio2024followers}
\begin{barticle}
\bauthor{\bsnm{Sangiorgio}, \binits{E.}},
\bauthor{\bsnm{Cinelli}, \binits{M.}},
\bauthor{\bsnm{Cerqueti}, \binits{R.}},
\bauthor{\bsnm{Quattrociocchi}, \binits{W.}}:
\batitle{Followers do not dictate the virality of news outlets on social media}.
\bjtitle{PNAS nexus}
\bvolume{3}(\bissue{7}),
\bfpage{257}
(\byear{2024})
\end{barticle}
\endbibitem

\bibitem[\protect\citeauthoryear{Etta et~al.}{2023}]{etta2023characterizing}
\begin{barticle}
\bauthor{\bsnm{Etta}, \binits{G.}},
\bauthor{\bsnm{Sangiorgio}, \binits{E.}},
\bauthor{\bsnm{Di~Marco}, \binits{N.}},
\bauthor{\bsnm{Avalle}, \binits{M.}},
\bauthor{\bsnm{Scala}, \binits{A.}},
\bauthor{\bsnm{Cinelli}, \binits{M.}},
\bauthor{\bsnm{Quattrociocchi}, \binits{W.}}:
\batitle{Characterizing engagement dynamics across topics on facebook}.
\bjtitle{Plos one}
\bvolume{18}(\bissue{6}),
\bfpage{0286150}
(\byear{2023})
\end{barticle}
\endbibitem

\bibitem[\protect\citeauthoryear{Zollo et~al.}{2026}]{zollo2026examining}
\begin{barticle}
\bauthor{\bsnm{Zollo}, \binits{S.}},
\bauthor{\bsnm{Sangiorgio}, \binits{E.}},
\bauthor{\bsnm{Etta}, \binits{G.}},
\bauthor{\bsnm{Di~Marco}, \binits{N.}},
\bauthor{\bsnm{Cinelli}, \binits{M.}},
\bauthor{\bsnm{Cerqueti}, \binits{R.}},
\bauthor{\bsnm{Quattrociocchi}, \binits{W.}}:
\batitle{Examining the relationship between content length and engagement with news outlets on multiple social media platforms}.
\bjtitle{Technological Forecasting and Social Change}
\bvolume{228},
\bfpage{124668}
(\byear{2026})
\end{barticle}
\endbibitem

\bibitem[\protect\citeauthoryear{Gallup}{2018}]{gallup2018indicators}
\begin{botherref}
\oauthor{\bsnm{Gallup}, \binits{K.}}:
Indicators of news media trust.
John S. and James L. Knight Foundation Miami
(2018)
\end{botherref}
\endbibitem

\bibitem[\protect\citeauthoryear{Nic et~al.}{2018}]{nic2018reuters}
\begin{botherref}
\oauthor{\bsnm{Nic}, \binits{N.}},
\oauthor{\bsnm{Fletcher}, \binits{R.}},
\oauthor{\bsnm{Kalogeropoulos}, \binits{A.}},
\oauthor{\bsnm{Levy}, \binits{D.A.}},
\oauthor{\bsnm{Nielsen}, \binits{R.K.}}:
Reuters institute digital news report 2018.
Reuters Institute for the Study of Journalism
\textbf{39}
(2018)
\end{botherref}
\endbibitem

\bibitem[\protect\citeauthoryear{Bessi et~al.}{2015}]{bessi2015science}
\begin{barticle}
\bauthor{\bsnm{Bessi}, \binits{A.}},
\bauthor{\bsnm{Coletto}, \binits{M.}},
\bauthor{\bsnm{Davidescu}, \binits{G.A.}},
\bauthor{\bsnm{Scala}, \binits{A.}},
\bauthor{\bsnm{Caldarelli}, \binits{G.}},
\bauthor{\bsnm{Quattrociocchi}, \binits{W.}}:
\batitle{Science vs conspiracy: Collective narratives in the age of misinformation}.
\bjtitle{PloS one}
\bvolume{10}(\bissue{2}),
\bfpage{0118093}
(\byear{2015})
\end{barticle}
\endbibitem

\bibitem[\protect\citeauthoryear{Cinelli et~al.}{2021}]{cinelli2021echo}
\begin{barticle}
\bauthor{\bsnm{Cinelli}, \binits{M.}},
\bauthor{\bsnm{De~Francisci~Morales}, \binits{G.}},
\bauthor{\bsnm{Galeazzi}, \binits{A.}},
\bauthor{\bsnm{Quattrociocchi}, \binits{W.}},
\bauthor{\bsnm{Starnini}, \binits{M.}}:
\batitle{The echo chamber effect on social media}.
\bjtitle{Proceedings of the National Academy of Sciences}
\bvolume{118}(\bissue{9}),
\bfpage{2023301118}
(\byear{2021})
\end{barticle}
\endbibitem

\bibitem[\protect\citeauthoryear{Terren and Borge-Bravo}{2021}]{terren2021echo}
\begin{botherref}
\oauthor{\bsnm{Terren}, \binits{L.T.L.}},
\oauthor{\bsnm{Borge-Bravo}, \binits{R.B.-B.R.}}:
Echo chambers on social media: A systematic review of the literature.
Review of Communication Research
\textbf{9}
(2021)
\end{botherref}
\endbibitem

\bibitem[\protect\citeauthoryear{Tucker et~al.}{2018}]{tucker2018social}
\begin{botherref}
\oauthor{\bsnm{Tucker}, \binits{J.A.}},
\oauthor{\bsnm{Guess}, \binits{A.}},
\oauthor{\bsnm{Barber{\'a}}, \binits{P.}},
\oauthor{\bsnm{Vaccari}, \binits{C.}},
\oauthor{\bsnm{Siegel}, \binits{A.}},
\oauthor{\bsnm{Sanovich}, \binits{S.}},
\oauthor{\bsnm{Stukal}, \binits{D.}},
\oauthor{\bsnm{Nyhan}, \binits{B.}}:
Social media, political polarization, and political disinformation: A review of the scientific literature.
Political polarization, and political disinformation: a review of the scientific literature (March 19, 2018)
(2018)
\end{botherref}
\endbibitem

\bibitem[\protect\citeauthoryear{Hartmann et~al.}{2025}]{hartmann2025systematic}
\begin{barticle}
\bauthor{\bsnm{Hartmann}, \binits{D.}},
\bauthor{\bsnm{Wang}, \binits{S.M.}},
\bauthor{\bsnm{Pohlmann}, \binits{L.}},
\bauthor{\bsnm{Berendt}, \binits{B.}}:
\batitle{A systematic review of echo chamber research: comparative analysis of conceptualizations, operationalizations, and varying outcomes}.
\bjtitle{Journal of Computational Social Science}
\bvolume{8}(\bissue{2}),
\bfpage{52}
(\byear{2025})
\end{barticle}
\endbibitem

\bibitem[\protect\citeauthoryear{Barber{\'a}}{2020}]{barbera2020social}
\begin{barticle}
\bauthor{\bsnm{Barber{\'a}}, \binits{P.}}:
\batitle{Social media, echo chambers, and political polarization}.
\bjtitle{Social media and democracy: The state of the field, prospects for reform}
\bvolume{34},
\bfpage{34}--\blpage{55}
(\byear{2020})
\end{barticle}
\endbibitem

\bibitem[\protect\citeauthoryear{Diaz~Ruiz and Nilsson}{2023}]{diaz2023disinformation}
\begin{barticle}
\bauthor{\bsnm{Diaz~Ruiz}, \binits{C.}},
\bauthor{\bsnm{Nilsson}, \binits{T.}}:
\batitle{Disinformation and echo chambers: how disinformation circulates on social media through identity-driven controversies}.
\bjtitle{Journal of public policy \& marketing}
\bvolume{42}(\bissue{1}),
\bfpage{18}--\blpage{35}
(\byear{2023})
\end{barticle}
\endbibitem

\bibitem[\protect\citeauthoryear{Bail et~al.}{2018}]{bail2018exposure}
\begin{barticle}
\bauthor{\bsnm{Bail}, \binits{C.A.}},
\bauthor{\bsnm{Argyle}, \binits{L.P.}},
\bauthor{\bsnm{Brown}, \binits{T.W.}},
\bauthor{\bsnm{Bumpus}, \binits{J.P.}},
\bauthor{\bsnm{Chen}, \binits{H.}},
\bauthor{\bsnm{Hunzaker}, \binits{M.F.}},
\bauthor{\bsnm{Lee}, \binits{J.}},
\bauthor{\bsnm{Mann}, \binits{M.}},
\bauthor{\bsnm{Merhout}, \binits{F.}},
\bauthor{\bsnm{Volfovsky}, \binits{A.}}:
\batitle{Exposure to opposing views on social media can increase political polarization}.
\bjtitle{Proceedings of the National Academy of Sciences}
\bvolume{115}(\bissue{37}),
\bfpage{9216}--\blpage{9221}
(\byear{2018})
\end{barticle}
\endbibitem

\bibitem[\protect\citeauthoryear{Kubin and Von~Sikorski}{2021}]{kubin2021role}
\begin{barticle}
\bauthor{\bsnm{Kubin}, \binits{E.}},
\bauthor{\bsnm{Von~Sikorski}, \binits{C.}}:
\batitle{The role of (social) media in political polarization: a systematic review}.
\bjtitle{Annals of the International Communication Association}
\bvolume{45}(\bissue{3}),
\bfpage{188}--\blpage{206}
(\byear{2021})
\end{barticle}
\endbibitem

\bibitem[\protect\citeauthoryear{Avalle et~al.}{2024}]{avalle2024persistent}
\begin{barticle}
\bauthor{\bsnm{Avalle}, \binits{M.}},
\bauthor{\bsnm{Di~Marco}, \binits{N.}},
\bauthor{\bsnm{Etta}, \binits{G.}},
\bauthor{\bsnm{Sangiorgio}, \binits{E.}},
\bauthor{\bsnm{Alipour}, \binits{S.}},
\bauthor{\bsnm{Bonetti}, \binits{A.}},
\bauthor{\bsnm{Alvisi}, \binits{L.}},
\bauthor{\bsnm{Scala}, \binits{A.}},
\bauthor{\bsnm{Baronchelli}, \binits{A.}},
\bauthor{\bsnm{Cinelli}, \binits{M.}}, \betal:
\batitle{Persistent interaction patterns across social media platforms and over time}.
\bjtitle{Nature}
\bvolume{628}(\bissue{8008}),
\bfpage{582}--\blpage{589}
(\byear{2024})
\end{barticle}
\endbibitem

\bibitem[\protect\citeauthoryear{Tahmasbi et~al.}{2021}]{tahmasbi2021go}
\begin{bchapter}
\bauthor{\bsnm{Tahmasbi}, \binits{F.}},
\bauthor{\bsnm{Schild}, \binits{L.}},
\bauthor{\bsnm{Ling}, \binits{C.}},
\bauthor{\bsnm{Blackburn}, \binits{J.}},
\bauthor{\bsnm{Stringhini}, \binits{G.}},
\bauthor{\bsnm{Zhang}, \binits{Y.}},
\bauthor{\bsnm{Zannettou}, \binits{S.}}:
\bctitle{“go eat a bat, chang!”: On the emergence of sinophobic behavior on web communities in the face of covid-19}.
In: \bbtitle{Proceedings of the Web Conference 2021},
pp. \bfpage{1122}--\blpage{1133}
(\byear{2021})
\end{bchapter}
\endbibitem

\bibitem[\protect\citeauthoryear{League}{2020}]{league2020online}
\begin{barticle}
\bauthor{\bsnm{League}, \binits{A.-D.}}:
\batitle{Online hate and harassment report: the american experience 2020}.
\bjtitle{Retrieved September}
\bvolume{25},
\bfpage{2021}
(\byear{2020})
\end{barticle}
\endbibitem

\bibitem[\protect\citeauthoryear{Newell et~al.}{2016}]{newell2016user}
\begin{bchapter}
\bauthor{\bsnm{Newell}, \binits{E.}},
\bauthor{\bsnm{Jurgens}, \binits{D.}},
\bauthor{\bsnm{Saleem}, \binits{H.}},
\bauthor{\bsnm{Vala}, \binits{H.}},
\bauthor{\bsnm{Sassine}, \binits{J.}},
\bauthor{\bsnm{Armstrong}, \binits{C.}},
\bauthor{\bsnm{Ruths}, \binits{D.}}:
\bctitle{User migration in online social networks: A case study on reddit during a period of community unrest}.
In: \bbtitle{Proceedings of the International AAAI Conference on Web and Social Media},
vol. \bseriesno{10},
pp. \bfpage{279}--\blpage{288}
(\byear{2016})
\end{bchapter}
\endbibitem

\bibitem[\protect\citeauthoryear{Horta~Ribeiro et~al.}{2021}]{horta2021platform}
\begin{barticle}
\bauthor{\bsnm{Horta~Ribeiro}, \binits{M.}},
\bauthor{\bsnm{Jhaver}, \binits{S.}},
\bauthor{\bsnm{Zannettou}, \binits{S.}},
\bauthor{\bsnm{Blackburn}, \binits{J.}},
\bauthor{\bsnm{Stringhini}, \binits{G.}},
\bauthor{\bsnm{De~Cristofaro}, \binits{E.}},
\bauthor{\bsnm{West}, \binits{R.}}:
\batitle{Do platform migrations compromise content moderation? evidence from r/the\_donald and r/incels}.
\bjtitle{Proceedings of the ACM on Human-Computer Interaction}
\bvolume{5}(\bissue{CSCW2}),
\bfpage{1}--\blpage{24}
(\byear{2021})
\end{barticle}
\endbibitem

\bibitem[\protect\citeauthoryear{Ali et~al.}{2021}]{ali2021understanding}
\begin{bchapter}
\bauthor{\bsnm{Ali}, \binits{S.}},
\bauthor{\bsnm{Saeed}, \binits{M.H.}},
\bauthor{\bsnm{Aldreabi}, \binits{E.}},
\bauthor{\bsnm{Blackburn}, \binits{J.}},
\bauthor{\bsnm{De~Cristofaro}, \binits{E.}},
\bauthor{\bsnm{Zannettou}, \binits{S.}},
\bauthor{\bsnm{Stringhini}, \binits{G.}}:
\bctitle{Understanding the effect of deplatforming on social networks}.
In: \bbtitle{Proceedings of the 13th ACM Web Science Conference 2021},
pp. \bfpage{187}--\blpage{195}
(\byear{2021})
\end{bchapter}
\endbibitem

\bibitem[\protect\citeauthoryear{Monti et~al.}{2023}]{monti2023online}
\begin{barticle}
\bauthor{\bsnm{Monti}, \binits{C.}},
\bauthor{\bsnm{Cinelli}, \binits{M.}},
\bauthor{\bsnm{Valensise}, \binits{C.}},
\bauthor{\bsnm{Quattrociocchi}, \binits{W.}},
\bauthor{\bsnm{Starnini}, \binits{M.}}:
\batitle{Online conspiracy communities are more resilient to deplatforming}.
\bjtitle{PNAS nexus}
\bvolume{2}(\bissue{10}),
\bfpage{324}
(\byear{2023})
\end{barticle}
\endbibitem

\bibitem[\protect\citeauthoryear{Schmitz and Samory}{2025}]{schmitz2025volunteerism}
\begin{barticle}
\bauthor{\bsnm{Schmitz}, \binits{A.}},
\bauthor{\bsnm{Samory}, \binits{M.}}:
\batitle{From volunteerism to corporatization: Analyzing participation in the 2015 and 2023 reddit blackouts}.
\bjtitle{Social Media+ Society}
\bvolume{11}(\bissue{1}),
\bfpage{20563051241309497}
(\byear{2025})
\end{barticle}
\endbibitem

\bibitem[\protect\citeauthoryear{Cava et~al.}{2023}]{cava2023drivers}
\begin{barticle}
\bauthor{\bsnm{Cava}, \binits{L.L.}},
\bauthor{\bsnm{Aiello}, \binits{L.M.}},
\bauthor{\bsnm{Tagarelli}, \binits{A.}}:
\batitle{Drivers of social influence in the twitter migration to mastodon}.
\bjtitle{Scientific Reports}
\bvolume{13}(\bissue{1}),
\bfpage{21626}
(\byear{2023})
\end{barticle}
\endbibitem

\bibitem[\protect\citeauthoryear{Failla and Rossetti}{2024}]{failla2024m}
\begin{barticle}
\bauthor{\bsnm{Failla}, \binits{A.}},
\bauthor{\bsnm{Rossetti}, \binits{G.}}:
\batitle{“i’m in the bluesky tonight”: Insights from a year worth of social data}.
\bjtitle{PloS one}
\bvolume{19}(\bissue{11}),
\bfpage{0310330}
(\byear{2024})
\end{barticle}
\endbibitem

\bibitem[\protect\citeauthoryear{Quelle and Bovet}{2025}]{quelle2025bluesky}
\begin{barticle}
\bauthor{\bsnm{Quelle}, \binits{D.}},
\bauthor{\bsnm{Bovet}, \binits{A.}}:
\batitle{Bluesky: Network topology, polarization, and algorithmic curation}.
\bjtitle{PloS one}
\bvolume{20}(\bissue{2}),
\bfpage{0318034}
(\byear{2025})
\end{barticle}
\endbibitem

\bibitem[\protect\citeauthoryear{Fiorina et~al.}{2004}]{fiorina2004culture}
\begin{botherref}
\oauthor{\bsnm{Fiorina}, \binits{M.P.}},
\oauthor{\bsnm{Abrams}, \binits{S.J.}},
\oauthor{\bsnm{Pope}, \binits{J.}}:
Culture war? the myth of a polarized america
(2004)
\end{botherref}
\endbibitem

\bibitem[\protect\citeauthoryear{Levendusky}{2009}]{levendusky2009partisan}
\begin{bbook}
\bauthor{\bsnm{Levendusky}, \binits{M.}}:
\bbtitle{The Partisan Sort: How Liberals Became Democrats and Conservatives Became Republicans},
(\byear{2009})
\end{bbook}
\endbibitem

\bibitem[\protect\citeauthoryear{Di~Martino et~al.}{2025}]{di2025ideological}
\begin{barticle}
\bauthor{\bsnm{Di~Martino}, \binits{E.}},
\bauthor{\bsnm{Galeazzi}, \binits{A.}},
\bauthor{\bsnm{Starnini}, \binits{M.}},
\bauthor{\bsnm{Quattrociocchi}, \binits{W.}},
\bauthor{\bsnm{Cinelli}, \binits{M.}}:
\batitle{Ideological fragmentation of the social media ecosystem: From echo chambers to echo platforms}.
\bjtitle{PNAS Nexus}
\bvolume{4}(\bissue{9}),
\bfpage{262}
(\byear{2025})
\end{barticle}
\endbibitem

\bibitem[\protect\citeauthoryear{Kupferschmidt}{2024}]{kupferschmidt2024like}
\begin{barticle}
\bauthor{\bsnm{Kupferschmidt}, \binits{K.}}:
\batitle{Like ‘old twitter’: the scientific community finds a new home on bluesky}.
\bjtitle{Science}
\bvolume{386},
\bfpage{950}--\blpage{6}
(\byear{2024})
\end{barticle}
\endbibitem

\bibitem[\protect\citeauthoryear{M.~Otala et~al.}{2021}]{m2021political}
\begin{bchapter}
\bauthor{\bsnm{M.~Otala}, \binits{J.}},
\bauthor{\bsnm{Kurtic}, \binits{G.}},
\bauthor{\bsnm{Grasso}, \binits{I.}},
\bauthor{\bsnm{Liu}, \binits{Y.}},
\bauthor{\bsnm{Matthews}, \binits{J.}},
\bauthor{\bsnm{Madraki}, \binits{G.}}:
\bctitle{Political polarization and platform migration: A study of parler and twitter usage by united states of america congress members}.
In: \bbtitle{Companion Proceedings of the Web Conference 2021},
pp. \bfpage{224}--\blpage{231}
(\byear{2021})
\end{bchapter}
\endbibitem

\bibitem[\protect\citeauthoryear{B{\"a}r et~al.}{2023}]{bar2023new}
\begin{barticle}
\bauthor{\bsnm{B{\"a}r}, \binits{D.}},
\bauthor{\bsnm{Pr{\"o}llochs}, \binits{N.}},
\bauthor{\bsnm{Feuerriegel}, \binits{S.}}:
\batitle{New threats to society from free-speech social media platforms}.
\bjtitle{Communications of the ACM}
\bvolume{66}(\bissue{10}),
\bfpage{37}--\blpage{40}
(\byear{2023})
\end{barticle}
\endbibitem

\bibitem[\protect\citeauthoryear{Stroud}{2010}]{stroud2010polarization}
\begin{barticle}
\bauthor{\bsnm{Stroud}, \binits{N.J.}}:
\batitle{Polarization and partisan selective exposure}.
\bjtitle{Journal of communication}
\bvolume{60}(\bissue{3}),
\bfpage{556}--\blpage{576}
(\byear{2010})
\end{barticle}
\endbibitem

\bibitem[\protect\citeauthoryear{Spohr}{2017}]{spohr2017fake}
\begin{barticle}
\bauthor{\bsnm{Spohr}, \binits{D.}}:
\batitle{Fake news and ideological polarization: Filter bubbles and selective exposure on social media}.
\bjtitle{Business information review}
\bvolume{34}(\bissue{3}),
\bfpage{150}--\blpage{160}
(\byear{2017})
\end{barticle}
\endbibitem

\bibitem[\protect\citeauthoryear{Brugnoli et~al.}{2019}]{brugnoli2019recursive}
\begin{barticle}
\bauthor{\bsnm{Brugnoli}, \binits{E.}},
\bauthor{\bsnm{Cinelli}, \binits{M.}},
\bauthor{\bsnm{Quattrociocchi}, \binits{W.}},
\bauthor{\bsnm{Scala}, \binits{A.}}:
\batitle{Recursive patterns in online echo chambers}.
\bjtitle{Scientific reports}
\bvolume{9}(\bissue{1}),
\bfpage{20118}
(\byear{2019})
\end{barticle}
\endbibitem

\bibitem[\protect\citeauthoryear{Cardenal et~al.}{2019}]{cardenal2019digital}
\begin{barticle}
\bauthor{\bsnm{Cardenal}, \binits{A.S.}},
\bauthor{\bsnm{Aguilar-Paredes}, \binits{C.}},
\bauthor{\bsnm{Galais}, \binits{C.}},
\bauthor{\bsnm{P{\'e}rez-Montoro}, \binits{M.}}:
\batitle{Digital technologies and selective exposure: How choice and filter bubbles shape news media exposure}.
\bjtitle{The international journal of press/politics}
\bvolume{24}(\bissue{4}),
\bfpage{465}--\blpage{486}
(\byear{2019})
\end{barticle}
\endbibitem

\bibitem[\protect\citeauthoryear{Pecile et~al.}{2025}]{pecile2025mapping}
\begin{barticle}
\bauthor{\bsnm{Pecile}, \binits{G.}},
\bauthor{\bsnm{Di~Marco}, \binits{N.}},
\bauthor{\bsnm{Cinelli}, \binits{M.}},
\bauthor{\bsnm{Quattrociocchi}, \binits{W.}}:
\batitle{Mapping the global election landscape on social media in 2024}.
\bjtitle{PloS one}
\bvolume{20}(\bissue{2}),
\bfpage{0316271}
(\byear{2025})
\end{barticle}
\endbibitem

\bibitem[\protect\citeauthoryear{Stefanov et~al.}{2020}]{stefanov2020predicting}
\begin{bchapter}
\bauthor{\bsnm{Stefanov}, \binits{P.}},
\bauthor{\bsnm{Darwish}, \binits{K.}},
\bauthor{\bsnm{Atanasov}, \binits{A.}},
\bauthor{\bsnm{Nakov}, \binits{P.}}:
\bctitle{Predicting the topical stance and political leaning of media using tweets}.
In: \bbtitle{Proceedings of the 58th Annual Meeting of the Association for Computational Linguistics},
pp. \bfpage{527}--\blpage{537}
(\byear{2020})
\end{bchapter}
\endbibitem

\bibitem[\protect\citeauthoryear{Flamino et~al.}{2023}]{flamino2023political}
\begin{barticle}
\bauthor{\bsnm{Flamino}, \binits{J.}},
\bauthor{\bsnm{Galeazzi}, \binits{A.}},
\bauthor{\bsnm{Feldman}, \binits{S.}},
\bauthor{\bsnm{Macy}, \binits{M.W.}},
\bauthor{\bsnm{Cross}, \binits{B.}},
\bauthor{\bsnm{Zhou}, \binits{Z.}},
\bauthor{\bsnm{Serafino}, \binits{M.}},
\bauthor{\bsnm{Bovet}, \binits{A.}},
\bauthor{\bsnm{Makse}, \binits{H.A.}},
\bauthor{\bsnm{Szymanski}, \binits{B.K.}}:
\batitle{Political polarization of news media and influencers on twitter in the 2016 and 2020 us presidential elections}.
\bjtitle{Nature Human Behaviour}
\bvolume{7}(\bissue{6}),
\bfpage{904}--\blpage{916}
(\byear{2023})
\end{barticle}
\endbibitem

\bibitem[\protect\citeauthoryear{Di~Marco et~al.}{2024}]{di2024users}
\begin{barticle}
\bauthor{\bsnm{Di~Marco}, \binits{N.}},
\bauthor{\bsnm{Cinelli}, \binits{M.}},
\bauthor{\bsnm{Alipour}, \binits{S.}},
\bauthor{\bsnm{Quattrociocchi}, \binits{W.}}:
\batitle{Users volatility on reddit and voat}.
\bjtitle{IEEE Transactions on Computational Social Systems}
\bvolume{11}(\bissue{5}),
\bfpage{5871}--\blpage{5879}
(\byear{2024})
\end{barticle}
\endbibitem

\bibitem[\protect\citeauthoryear{Dehghan and Nagappa}{2022}]{dehghan2022politicization}
\begin{barticle}
\bauthor{\bsnm{Dehghan}, \binits{E.}},
\bauthor{\bsnm{Nagappa}, \binits{A.}}:
\batitle{Politicization and radicalization of discourses in the alt-tech ecosystem: A case study on gab social}.
\bjtitle{Social Media+ Society}
\bvolume{8}(\bissue{3}),
\bfpage{20563051221113075}
(\byear{2022})
\end{barticle}
\endbibitem

\bibitem[\protect\citeauthoryear{Piao et~al.}{2023}]{piao2023human}
\begin{barticle}
\bauthor{\bsnm{Piao}, \binits{J.}},
\bauthor{\bsnm{Liu}, \binits{J.}},
\bauthor{\bsnm{Zhang}, \binits{F.}},
\bauthor{\bsnm{Su}, \binits{J.}},
\bauthor{\bsnm{Li}, \binits{Y.}}:
\batitle{Human--ai adaptive dynamics drives the emergence of information cocoons}.
\bjtitle{Nature Machine Intelligence}
\bvolume{5}(\bissue{11}),
\bfpage{1214}--\blpage{1224}
(\byear{2023})
\end{barticle}
\endbibitem

\bibitem[\protect\citeauthoryear{Schroeder et~al.}{2026}]{schroeder2026malicious}
\begin{barticle}
\bauthor{\bsnm{Schroeder}, \binits{D.T.}},
\bauthor{\bsnm{Cha}, \binits{M.}},
\bauthor{\bsnm{Baronchelli}, \binits{A.}},
\bauthor{\bsnm{Bostrom}, \binits{N.}},
\bauthor{\bsnm{Christakis}, \binits{N.A.}},
\bauthor{\bsnm{Garcia}, \binits{D.}},
\bauthor{\bsnm{Goldenberg}, \binits{A.}},
\bauthor{\bsnm{Kyrychenko}, \binits{Y.}},
\bauthor{\bsnm{Leyton-Brown}, \binits{K.}},
\bauthor{\bsnm{Lutz}, \binits{N.}}, \betal:
\batitle{How malicious ai swarms can threaten democracy}.
\bjtitle{Science}
\bvolume{391}(\bissue{6783}),
\bfpage{354}--\blpage{357}
(\byear{2026})
\end{barticle}
\endbibitem

\bibitem[\protect\citeauthoryear{Baumgartner et~al.}{2020}]{baumgartner2020pushshift}
\begin{bchapter}
\bauthor{\bsnm{Baumgartner}, \binits{J.}},
\bauthor{\bsnm{Zannettou}, \binits{S.}},
\bauthor{\bsnm{Keegan}, \binits{B.}},
\bauthor{\bsnm{Squire}, \binits{M.}},
\bauthor{\bsnm{Blackburn}, \binits{J.}}:
\bctitle{The pushshift reddit dataset}.
In: \bbtitle{Proceedings of the International AAAI Conference on Web and Social Media},
vol. \bseriesno{14},
pp. \bfpage{830}--\blpage{839}
(\byear{2020})
\end{bchapter}
\endbibitem

\bibitem[\protect\citeauthoryear{Balasubramanian et~al.}{2024}]{balasubramanian2024public}
\begin{botherref}
\oauthor{\bsnm{Balasubramanian}, \binits{A.}},
\oauthor{\bsnm{Zou}, \binits{V.}},
\oauthor{\bsnm{Narayana}, \binits{H.}},
\oauthor{\bsnm{You}, \binits{C.}},
\oauthor{\bsnm{Luceri}, \binits{L.}},
\oauthor{\bsnm{Ferrara}, \binits{E.}}:
A public dataset tracking social media discourse about the 2024 us presidential election on twitter/x.
arXiv preprint arXiv:2411.00376
(2024)
\end{botherref}
\endbibitem

\bibitem[\protect\citeauthoryear{Shah et~al.}{2024}]{shah2024unfiltered}
\begin{botherref}
\oauthor{\bsnm{Shah}, \binits{K.}},
\oauthor{\bsnm{Gerard}, \binits{P.}},
\oauthor{\bsnm{Luceri}, \binits{L.}},
\oauthor{\bsnm{Ferrara}, \binits{E.}}:
Unfiltered conversations: A dataset of 2024 us presidential election discourse on truth social.
arXiv preprint arXiv:2411.01330
(2024)
\end{botherref}
\endbibitem

\end{thebibliography}

\clearpage

\newcommand{\beginsupplement}{
    \setcounter{section}{0}
    \renewcommand{\thesection}{S\arabic{section}}
    \setcounter{equation}{0}
    \renewcommand{\theequation}{S\arabic{equation}}
    \setcounter{table}{0}
    \renewcommand{\thetable}{S\arabic{table}}
    \setcounter{figure}{0}
    \renewcommand{\thefigure}{S\arabic{figure}}
    \newcounter{SIfig}
    \renewcommand{\theSIfig}{S\arabic{SIfig}}}

\beginsupplement

\onecolumn
\section*{\centering Supplementary Information}

\subsection*{Power-law distribution of engagement metrics}

\begin{figure}[h]
    \centering
    \includegraphics[width=\columnwidth]{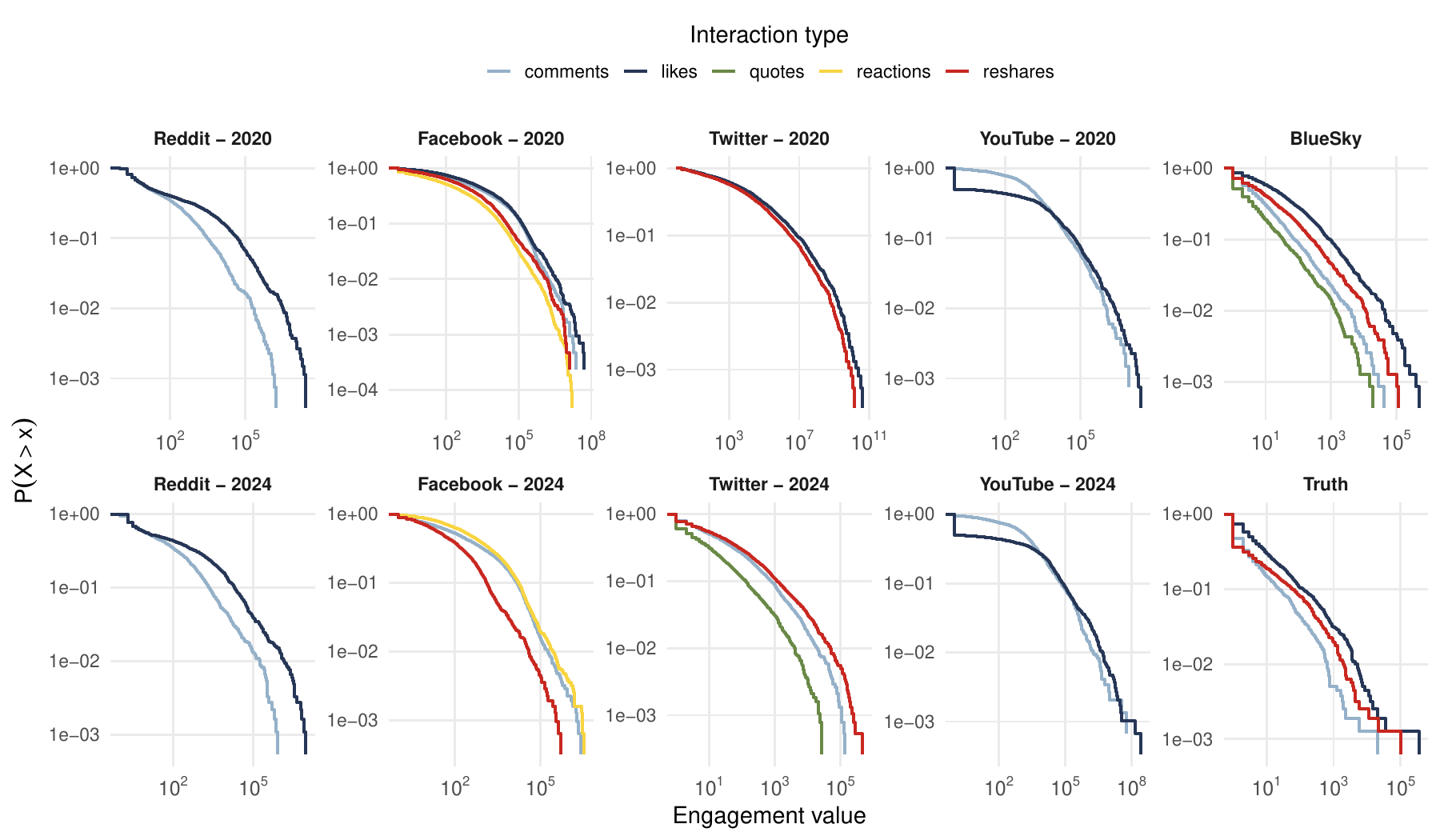}
    \caption{\textbf{Empirical complementary cumulative distribution function of engagement received by sources on the different platforms, divided by engagement metric}}
    \label{fig:powerlaw_engagement}
\end{figure}

Engagement is measured using platform-specific metrics, reflecting both structural differences across platforms and data availability constraints. Despite this heterogeneity, Figure~\ref{fig:powerlaw_engagement} shows that all engagement metrics exhibit similarly heavy-tailed distributions. This common structure supports cross-platform comparisons, as our analysis does not rely on raw engagement values, but on the fraction of total engagement received by different sources. Note how in the figure, for visual interpretability, the different interaction metrics were mapped onto five broader conceptual categories of engagement; for example, Reddit's ``score'' was grouped under the category of ``likes''. 

\newpage

\subsection*{Raw volume of news sharing patterns}
\begin{figure}[h]
    \centering
    \includegraphics[width=\columnwidth]{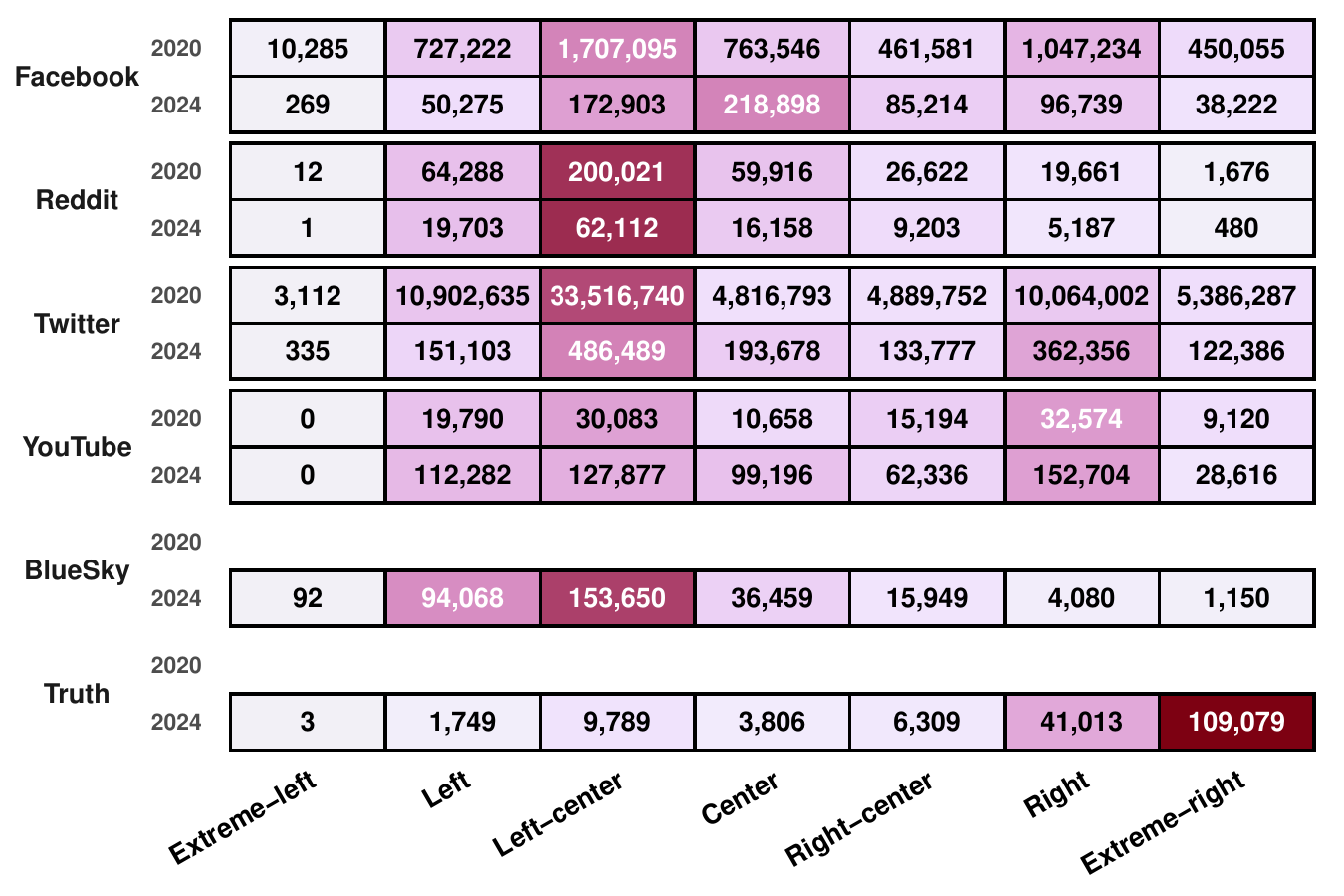}
    \caption{\textbf{Platforms’ news sharing patterns.} Volume of shared URLs linking to
            external news domains across platforms, stratified by political leaning according to
            MBFC labels.}
    \label{fig:diet_raw}
\end{figure}

We show in Figure~\ref{fig:diet_raw} the same results obtained in Figure~\ref{fig:weighted_diet} in the main manuscript, this time presenting raw volume instead of fraction of URLs shared. We also explicitly show the volume of domains classified as Extreme-left, without collapsing them into the Left category, as done in the main manuscript. Table~\ref{tab:delta_leanings} shows the change in number of URLs for sources of different leanings between the two electoral cycles for the four platforms with longitudinal data available.

\begin{table}[b]
\centering
\caption{Percentage change in total URLs between 2020 and 2024, stratified by leaning}
\centering
\begin{tabular}[t]{lrrrrrrr}
\toprule
Platform & Extreme-left & Left & Left-center & Center & Right-center & Right & Extreme-right\\
\midrule
Facebook & -97.4\% & -93.1\% & -89.9\% & -71.3\% & -81.5\% & -90.8\% & -91.5\%\\
Reddit & -91.7\% & -69.4\% & -68.9\% & -73.0\% & -65.4\% & -73.6\% & -71.4\%\\
Twitter & -89.2\% & -98.6\% & -98.5\% & -96.0\% & -97.3\% & -96.4\% & -97.7\%\\
YouTube & 0\% & 467.4\% & 325.1\% & 830.7\% & 310.3\% & 368.8\% & 213.8\%\\
\bottomrule
\end{tabular}
\label{tab:delta_leanings}
\end{table}

\newpage
\subsection*{Engagement towards questionable sources}

\begin{figure}[h]
    \centering
    \includegraphics[width=.8\columnwidth]{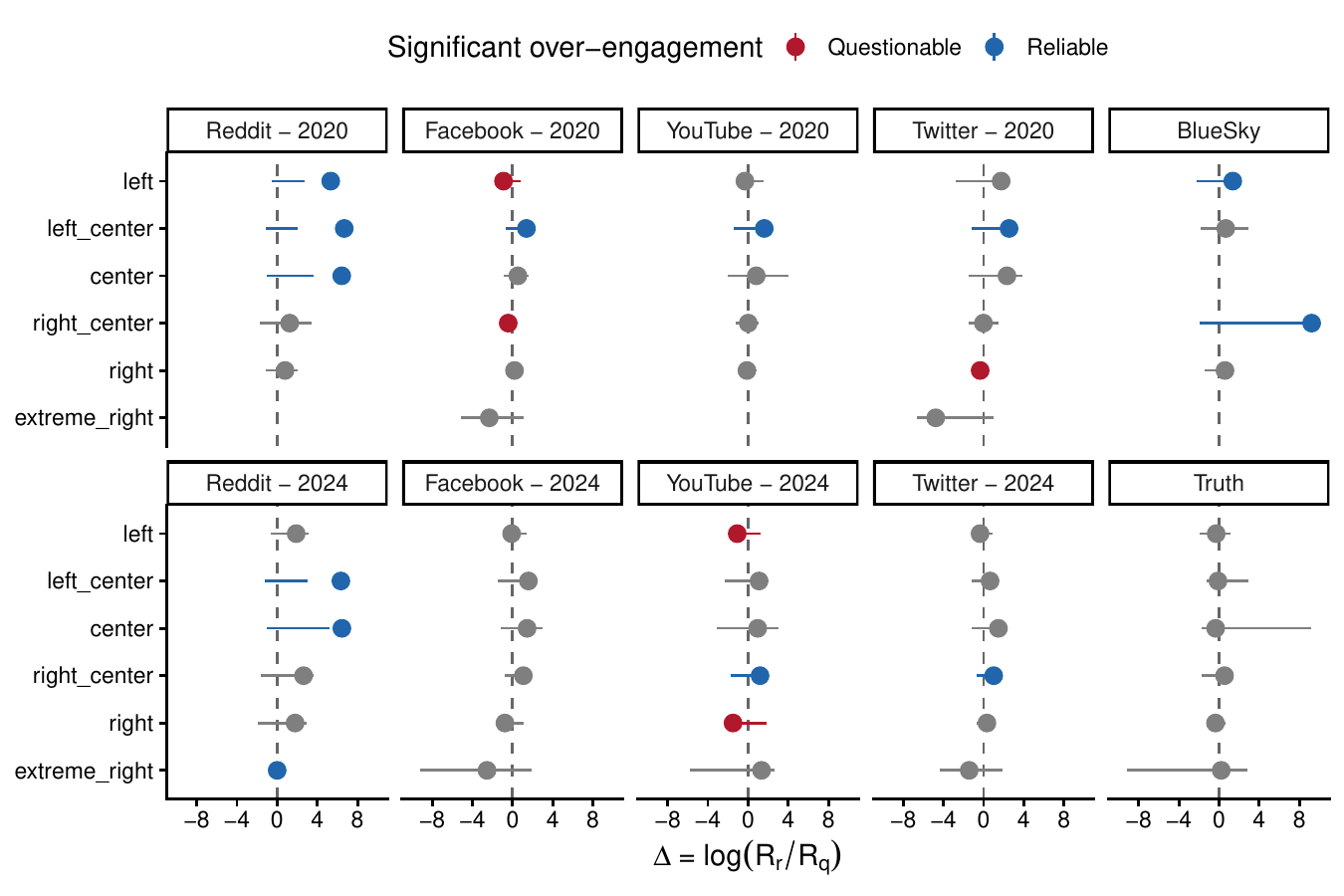}
    \caption{\textbf{Observed log differences in engagement between reliable and questionable sources across platforms and ideological leanings}. Dots represent the observed statistic
    $\Delta = \log(R_r / R_q)$. Horizontal lines indicate the central 95\% interval of the permutation-based null distribution. The dashed horizontal line at $\Delta = 0$ denotes equal relative engagement. Colors indicate whether the observed $\Delta$ reflects significantly greater engagement with reliable sources or significantly greater engagement with questionable sources.}
    \label{fig:engagement_quest}
\end{figure}

Figure \ref{fig:engagement_quest} show the full results of the second null model cited in Figure \ref{fig:randomization_ratio} of the main manuscript. With this procedure, we test the hypothesis that, conditional on platform, ideological leaning, and total engagement and shares, the assignment of engagement to questionable or reliable sources is random (i.e., it does not follow preferential behaviour by the users). To quantify whether engagement with questionable (or reliable) sources deviates from what would be expected under random allocation, we define for each platform and source leaning a log-difference quantity
\[
\Delta = \log\left(\frac{R_r}{R_q}\right) = \log(R_r) - \log(R_q),
\]
where $R_r$ and $R_q$ denote the engagement-to-supply ratios for reliable and questionable sources for a given platform and source leaning, respectively. Positive values of $\Delta$ indicate relatively greater engagement with reliable sources, while negative values indicate relatively greater engagement with questionable sources, with respect to baseline expectations. Significance is assessed using a permutation-based null model in which the reliability labels of sources are randomly reassigned within each platform and ideological bin, while preserving total engagement and supply. For each permutation, $\Delta$ is recomputed, yielding an empirical null distribution representing the values expected if engagement were independent of source reliability. A directional p-value is computed as the proportion of null realizations that are at least as extreme as the observed $\Delta$ in the direction of the effect. Small p-values indicate unusually high engagement with reliable sources, while large p-values (close to one) indicate unusually high engagement with questionable sources.

\newpage

\subsection*{Union set of top 20 domains by engagement}

\setlength{\tabcolsep}{3pt}
\renewcommand{\arraystretch}{0.9}
\begin{longtable}[h]{@{}lcc|cc|cc|cc|c|c@{}}
\caption{Percentage of engagement for the union set of top domains across platforms} \\
\label{tab:union_set} \\
\toprule
\multicolumn{1}{l}{ } & \multicolumn{2}{c}{Facebook} & \multicolumn{2}{c}{Reddit} & \multicolumn{2}{c}{YouTube} & \multicolumn{2}{c}{Twitter} & \multicolumn{1}{c}{Bluesky} & \multicolumn{1}{c}{Truth} \\
\cmidrule(l{3pt}r{3pt}){2-3} \cmidrule(l{3pt}r{3pt}){4-5} \cmidrule(l{3pt}r{3pt}){6-7} \cmidrule(l{3pt}r{3pt}){8-9} \cmidrule(l{3pt}c{3pt}){10-10} \cmidrule(l{3pt}c{3pt}){11-11}
Source & 2020 &2024 & 2020 & 2024 & 2020 & 2024 & 2020 & 2024 & 2024 & 2024\\
\midrule
abcnews.go.com & 1.2\% & 2.7\% & 0\% & 0\% & 0.7\% & 1.4\% & 0\% & 0\% & 0\% & 0\%\\
actblue.com & 0\% & 0\% & 0\% & 0\% & 1.4\% & 0\% & 11.8\% & 1.7\% & 0\% & 0\%\\
apnews.com & 0\% & 0\% & 0\% & 2.3\% & 0\% & 0\% & 1.6\% & 2.9\% & 1.8\% & 0\%\\
axios.com & 0\% & 0.8\% & 1.7\% & 1.5\% & 0\% & 0\% & 0\% & 1.5\% & 1.3\% & 0\%\\
bitchute.com & 0\% & 0\% & 0\% & 0\% & 1\% & 0.7\% & 0\% & 0\% & 0\% & 0.8\%\\
\addlinespace
bongino.com & 1.4\% & 0.9\% & 0\% & 0\% & 0\% & 0\% & 0\% & 0\% & 0\% & 0\%\\
breitbart.com & 3.2\% & 5.3\% & 0\% & 0\% & 0\% & 0\% & 3.1\% & 1.6\% & 0\% & 1.5\%\\
briantylercohen.com & 0\% & 0\% & 0\% & 0\% & 1\% & 1.5\% & 0\% & 0\% & 0\% & 0\%\\
businessinsider.com & 1\% & 0\% & 7.4\% & 1.9\% & 0\% & 0\% & 0\% & 0\% & 0\% & 0\%\\
cbsnews.com & 1.1\% & 0\% & 0\% & 0\% & 1.1\% & 1\% & 1.2\% & 0\% & 1.4\% & 0\%\\
\addlinespace
cnbc.com & 0\% & 0\% & 1.4\% & 0\% & 0.9\% & 0\% & 1.2\% & 0\% & 0\% & 0\%\\
cnn.com & 4.1\% & 4.2\% & 4.7\% & 2.9\% & 0\% & 0.9\% & 5.6\% & 3.3\% & 4.1\% & 0\%\\
commondreams.org & 0\% & 0\% & 4.8\% & 0\% & 0\% & 0\% & 0\% & 0\% & 0\% & 0\%\\
dailycaller.com & 0\% & 1.3\% & 0\% & 0\% & 0\% & 0\% & 0\% & 0\% & 0\% & 0\%\\
dailymail.co.uk & 0\% & 0\% & 0\% & 0\% & 0\% & 0.9\% & 0\% & 0\% & 0\% & 0\%\\
\addlinespace
dailywire.com & 5.7\% & 0\% & 0\% & 0\% & 0.7\% & 1\% & 0\% & 0\% & 0\% & 0\%\\
davidpakman.com & 0\% & 0\% & 0\% & 0\% & 0.7\% & 2\% & 0\% & 0\% & 0\% & 0\%\\
democracydocket.com & 0\% & 0\% & 0\% & 0\% & 0\% & 0\% & 0\% & 0\% & 5.3\% & 0\%\\
donaldjtrump.com & 0\% & 0\% & 0\% & 0\% & 0\% & 0\% & 0\% & 5.1\% & 0\% & 3.6\%\\
euromaidanpress.com & 0\% & 0\% & 0\% & 0\% & 0\% & 0.6\% & 0\% & 0\% & 0\% & 0\%\\
\addlinespace
foxbusiness.com & 0\% & 0\% & 0\% & 0\% & 2.7\% & 1.1\% & 0\% & 0\% & 0\% & 0\%\\
foxnews.com & 5.8\% & 6.3\% & 0\% & 0\% & 49.8\% & 31.7\% & 4\% & 3.2\% & 0\% & 2\%\\
frankspeech.com & 0\% & 0\% & 0\% & 0\% & 0\% & 0\% & 0\% & 0\% & 0\% & 1.8\%\\
hannity.com & 0\% & 2.7\% & 0\% & 0\% & 0\% & 0\% & 0\% & 0\% & 0\% & 0\%\\
huffpost.com & 2\% & 1.2\% & 3.4\% & 3.4\% & 0\% & 0\% & 0\% & 0\% & 0\% & 0\%\\
\addlinespace
justthenews.com & 0\% & 0\% & 0\% & 0\% & 0\% & 0\% & 0\% & 1.8\% & 0\% & 0.8\%\\
medium.com & 0\% & 0\% & 0\% & 0\% & 0\% & 0\% & 1.4\% & 0\% & 0\% & 0\%\\
meidasnews.com & 0\% & 0\% & 0\% & 0\% & 0\% & 0\% & 0\% & 2.3\% & 0\% & 0\%\\
motherjones.com & 0\% & 0\% & 1.5\% & 0\% & 0\% & 0\% & 0\% & 0\% & 0\% & 0\%\\
msnbc.com & 2.7\% & 1.4\% & 1.9\% & 3.5\% & 4.1\% & 27.6\% & 0\% & 0\% & 1.8\% & 0\%\\
\addlinespace
nbcnews.com & 3.1\% & 1.3\% & 1.8\% & 3.7\% & 2\% & 1.7\% & 2.2\% & 0\% & 4.2\% & 0\%\\
newrepublic.com & 0\% & 0\% & 0\% & 7.9\% & 0\% & 0\% & 0\% & 0\% & 4.4\% & 0\%\\
news.yahoo.com & 0.9\% & 0\% & 0\% & 0\% & 0\% & 0\% & 0\% & 0\% & 0\% & 0\%\\
newsbusters.org & 0\% & 0.8\% & 0\% & 0\% & 0\% & 0\% & 0\% & 0\% & 0\% & 0\%\\
newsmax.com & 0\% & 6\% & 0\% & 0\% & 0\% & 0\% & 0\% & 1.3\% & 0\% & 0.8\%\\
\addlinespace
newsnationnow.com & 0\% & 0\% & 0\% & 0\% & 0\% & 0.6\% & 0\% & 0\% & 0\% & 0\%\\
newsweek.com & 0\% & 0\% & 6.4\% & 9.6\% & 0\% & 0\% & 0\% & 3\% & 1.1\% & 0\%\\
nj.com & 0\% & 0\% & 0\% & 1.5\% & 0\% & 0\% & 0\% & 0\% & 0\% & 0\%\\
ntd.com & 0\% & 0\% & 0\% & 0\% & 0.8\% & 0\% & 0\% & 0\% & 0\% & 0.8\%\\
nypost.com & 0\% & 0\% & 0\% & 0\% & 0\% & 0\% & 3.7\% & 2.3\% & 0\% & 1.3\%\\
\addlinespace
nytimes.com & 2.4\% & 2.6\% & 2.2\% & 2.6\% & 0\% & 0\% & 10.4\% & 1.8\% & 11.4\% & 0\%\\
odysee.com & 0\% & 0\% & 0\% & 0\% & 0\% & 0.6\% & 0\% & 0\% & 0\% & 0\%\\
parler.com & 0\% & 0\% & 0\% & 0\% & 2.2\% & 0\% & 0\% & 0\% & 0\% & 0\%\\
politico.com & 0\% & 0\% & 1.8\% & 1.4\% & 0\% & 0\% & 2.1\% & 2.3\% & 2.6\% & 1\%\\
rawstory.com & 0\% & 0\% & 0\% & 0\% & 0\% & 0\% & 0\% & 2.1\% & 3.6\% & 0\%\\
\addlinespace
realclearpolitics.com & 0\% & 0\% & 0\% & 0\% & 0\% & 0\% & 0\% & 0\% & 0\% & 0.8\%\\
reuters.com & 0\% & 0\% & 0\% & 1.2\% & 0\% & 0\% & 0\% & 0\% & 1.9\% & 0\%\\
rollingstone.com & 0\% & 0\% & 0\% & 3.7\% & 0\% & 0\% & 0\% & 0\% & 1.1\% & 0\%\\
rumble.com & 0\% & 4.7\% & 0\% & 0\% & 0\% & 1.8\% & 0\% & 1.8\% & 0\% & 51.1\%\\
rvat.org & 0\% & 0\% & 0\% & 0\% & 0.7\% & 0\% & 1.6\% & 0\% & 0\% & 0\%\\
\addlinespace
salon.com & 0\% & 0\% & 3.1\% & 4.5\% & 0\% & 0\% & 0\% & 0\% & 1.1\% & 0\%\\
scientificamerican.com & 0\% & 0\% & 0\% & 0\% & 0\% & 0\% & 2.6\% & 0\% & 0\% & 0\%\\
sebgorka.com & 0\% & 0\% & 0\% & 0\% & 0\% & 0\% & 0\% & 0\% & 0\% & 1\%\\
slate.com & 0\% & 0\% & 1.4\% & 0\% & 0\% & 0\% & 0\% & 0\% & 0\% & 0\%\\
talkingpointsmemo.com & 0\% & 0\% & 1.3\% & 0\% & 0\% & 0\% & 0\% & 0\% & 0\% & 0\%\\
\addlinespace
the-sun.com & 0.9\% & 0\% & 0\% & 0\% & 0\% & 0\% & 0\% & 0\% & 0\% & 0\%\\
theatlantic.com & 0\% & 0\% & 0\% & 0\% & 0\% & 0\% & 2.3\% & 0\% & 2.3\% & 0\%\\
theblaze.com & 1.8\% & 2.3\% & 0\% & 0\% & 0.8\% & 0\% & 0\% & 0\% & 0\% & 0\%\\
thebulwark.com & 0\% & 0\% & 0\% & 0\% & 0\% & 1.2\% & 0\% & 0\% & 1.4\% & 0\%\\
thedailybeast.com & 0\% & 0\% & 3.3\% & 6.1\% & 0\% & 0\% & 1.4\% & 0\% & 0\% & 0\%\\
\addlinespace
theepochtimes.com & 0\% & 0\% & 0\% & 0\% & 0\% & 0\% & 0\% & 0\% & 0\% & 0.6\%\\
thefederalist.com & 0\% & 0\% & 0\% & 0\% & 0\% & 0\% & 1.2\% & 0\% & 0\% & 1\%\\
thegatewaypundit.com & 0\% & 0\% & 0\% & 0\% & 0\% & 0\% & 0\% & 1.7\% & 0\% & 6.5\%\\
theguardian.com & 0\% & 0\% & 1.9\% & 3.1\% & 0\% & 0\% & 0\% & 1.6\% & 3.6\% & 0\%\\
thehill.com & 3.1\% & 3.9\% & 7.8\% & 6.9\% & 0\% & 0\% & 1.4\% & 1.6\% & 1.1\% & 0\%\\
\addlinespace
thepostmillennial.com & 0\% & 0\% & 0\% & 0\% & 0\% & 0\% & 0\% & 0\% & 0\% & 0.6\%\\
thesun.co.uk & 0\% & 0\% & 0\% & 0\% & 0\% & 0.7\% & 0\% & 0\% & 0\% & 0\%\\
timcast.com & 0\% & 0\% & 0\% & 0\% & 0.8\% & 0.7\% & 0\% & 0\% & 0\% & 0\%\\
townhall.com & 0\% & 0\% & 0\% & 0\% & 0\% & 0\% & 0\% & 0\% & 0\% & 0.6\%\\
trendingpolitics.com & 1.6\% & 0\% & 0\% & 0\% & 0\% & 0\% & 0\% & 0\% & 0\% & 0\%\\
\addlinespace
trofire.com & 0\% & 0\% & 0\% & 0\% & 0.9\% & 0\% & 0\% & 0\% & 0\% & 0\%\\
turleytalks.com & 0\% & 0\% & 0\% & 0\% & 2.8\% & 0\% & 0\% & 0\% & 0\% & 0\%\\
tyt.com & 0\% & 0\% & 0\% & 0\% & 5\% & 3.8\% & 0\% & 0\% & 0\% & 0\%\\
usatoday.com & 0\% & 0\% & 1.4\% & 1.8\% & 0\% & 0\% & 2.3\% & 0\% & 0\% & 0\%\\
washingtonpost.com & 2.3\% & 1.9\% & 5.7\% & 2.5\% & 0\% & 0\% & 7.3\% & 1.2\% & 9\% & 0\%\\
\addlinespace
washingtontimes.com & 0.9\% & 0\% & 0\% & 0\% & 0\% & 0\% & 0\% & 0\% & 0\% & 0\%\\
westernjournal.com & 2.3\% & 1.6\% & 0\% & 0\% & 0\% & 0\% & 0\% & 0\% & 0\% & 0.8\%\\
wsj.com & 0\% & 1\% & 0\% & 0\% & 0\% & 0\% & 0\% & 0\% & 0\% & 0\%\\
yournews.com & 0\% & 0\% & 0\% & 0\% & 0\% & 0\% & 0\% & 0\% & 0\% & 2.9\%\\
\bottomrule
\end{longtable}

\newpage

\subsection*{Cosine similarity of engagement with different thresholds}

\begin{figure}[h]
    \centering
    \includegraphics[width=\columnwidth]{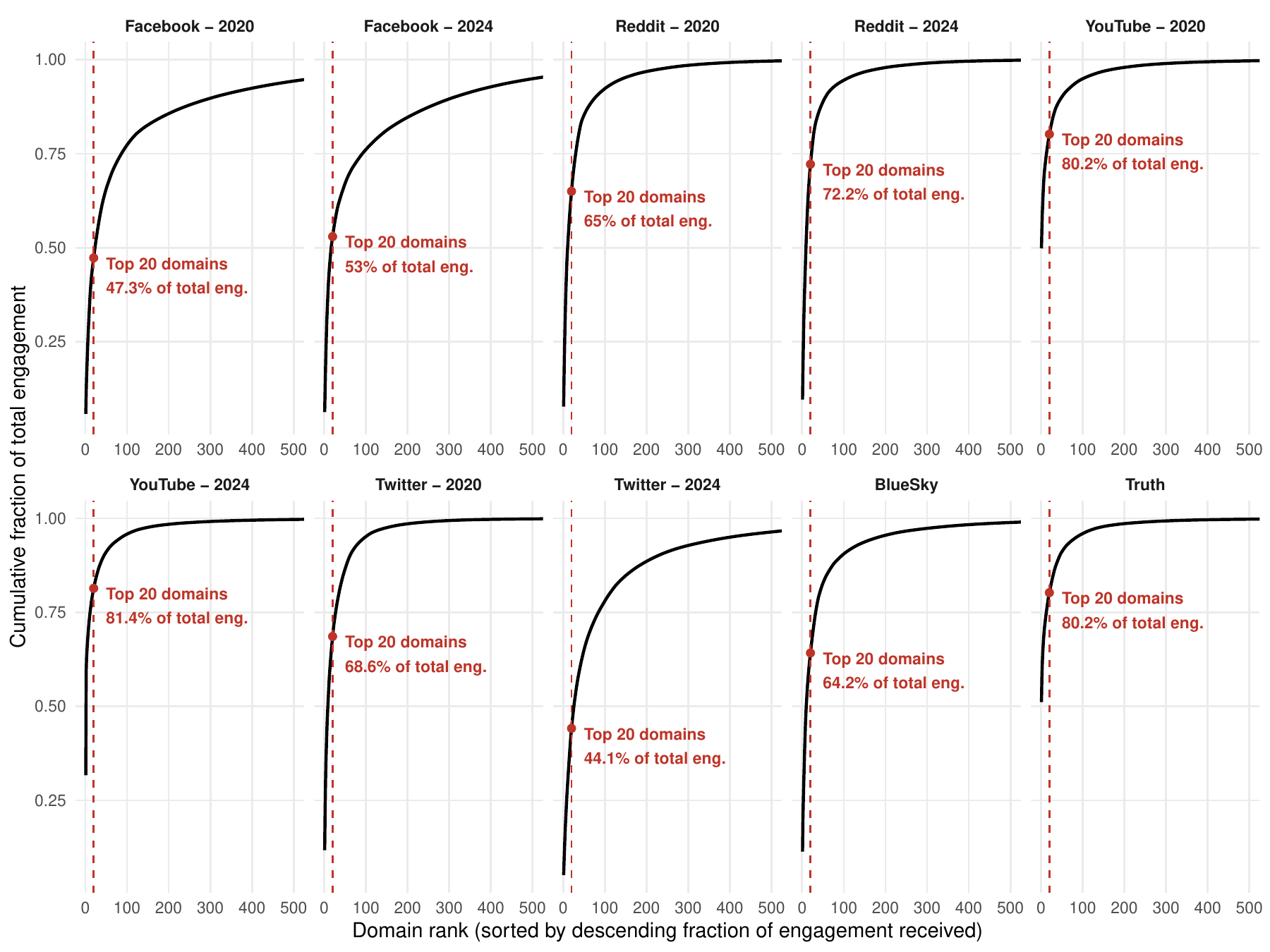}
    \caption{\textbf{Cumulative fraction of engagement received by different sources.}}
    \label{fig:engagement_distribution}
\end{figure}

\begin{figure}[h]
    \centering
    \includegraphics[width=\columnwidth]{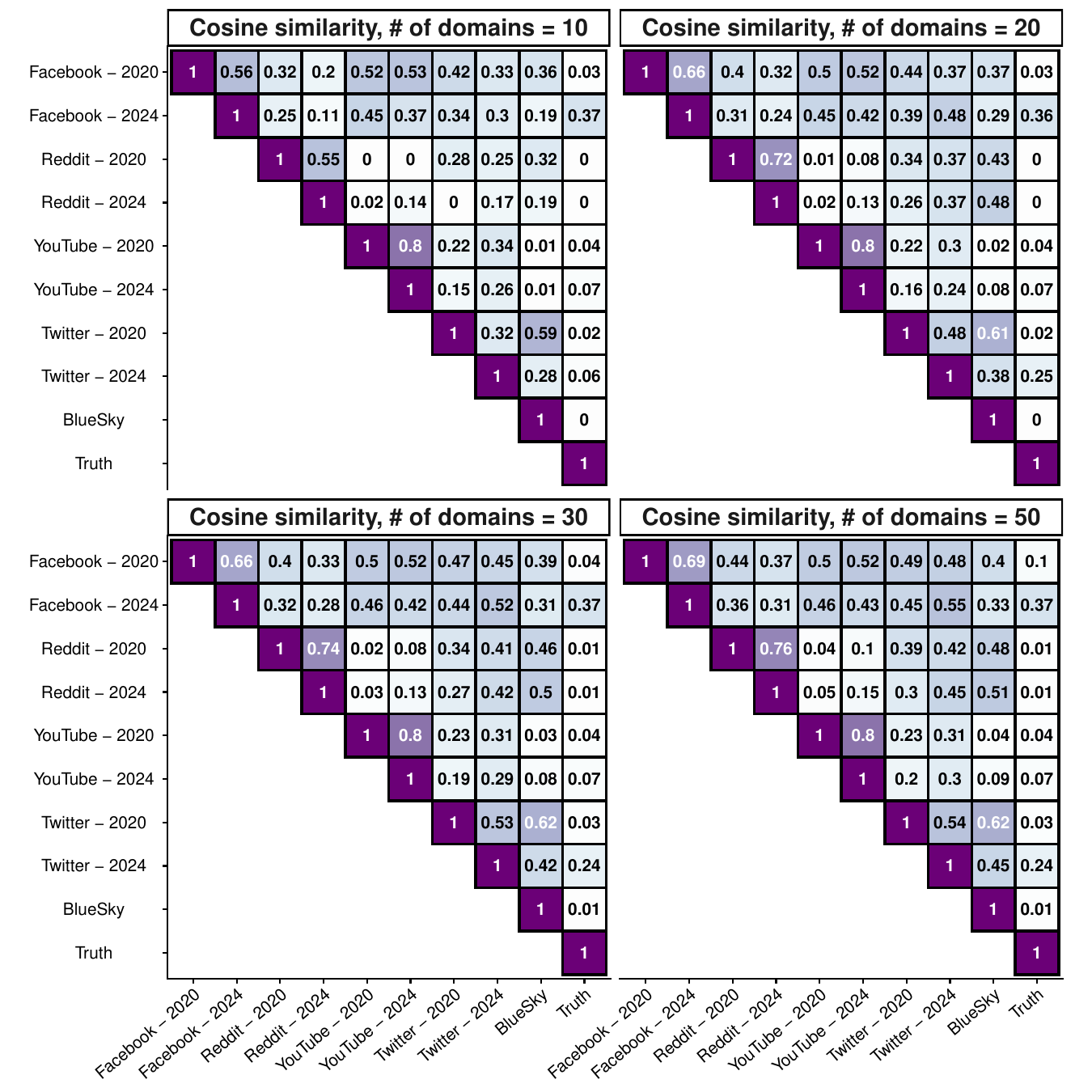}
    \caption{\textbf{Cosine similarity of engagement across platforms, considering the top 10, 20, 30, and 50 sources most engaged with.}}
    \label{fig:cosine_robustness}
\end{figure}

In the main analysis, cosine similarity between engagement patterns across platforms and time periods (Figure~\ref{fig:similarity_multiplex}) is computed using the top 20 most engaged domains. While this choice is necessarily somewhat arbitrary, it is motivated by the highly skewed distribution of engagement: as shown in Figure~\ref{fig:engagement_distribution}, a small number of domains accounts for the vast majority of interactions on each platform. To assess robustness, we replicate the analysis using alternative thresholds (Figure~\ref{fig:cosine_robustness}), showing that the results remain consistent, indicating that our findings are not driven by the specific threshold choice.

\newpage

\newpage

\subsection*{Common time window}

\begin{figure}[h]
    \centering
    \includegraphics[width=.75\columnwidth]{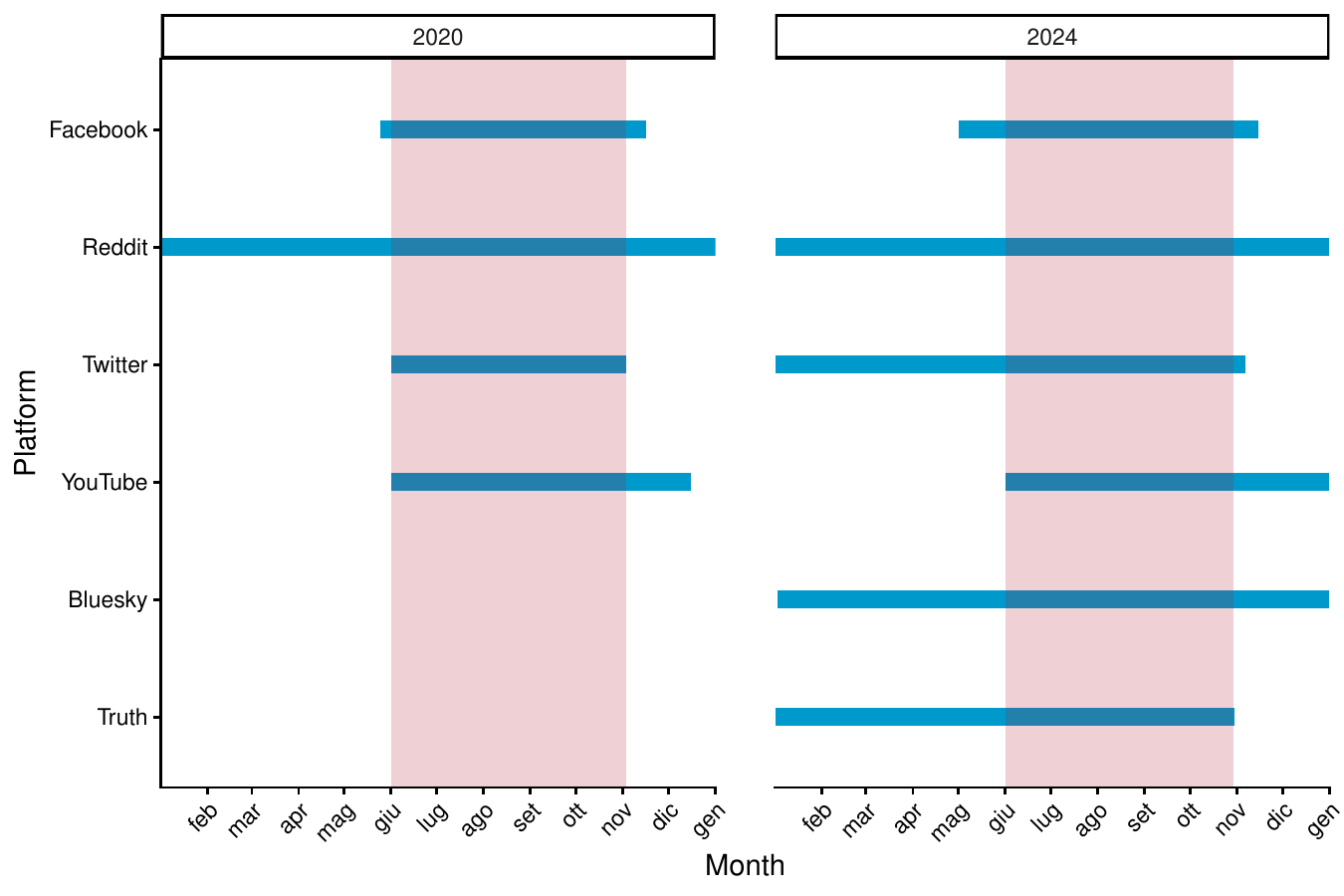}
    \caption{\textbf{Temporal coverage of the different data sets, and highlighted common time window (in red), differentiated by 2020 and 2024.}}
    \label{fig:common_windows}
\end{figure}

\begin{figure}[h]
    \centering
    \includegraphics[width=.75\columnwidth]{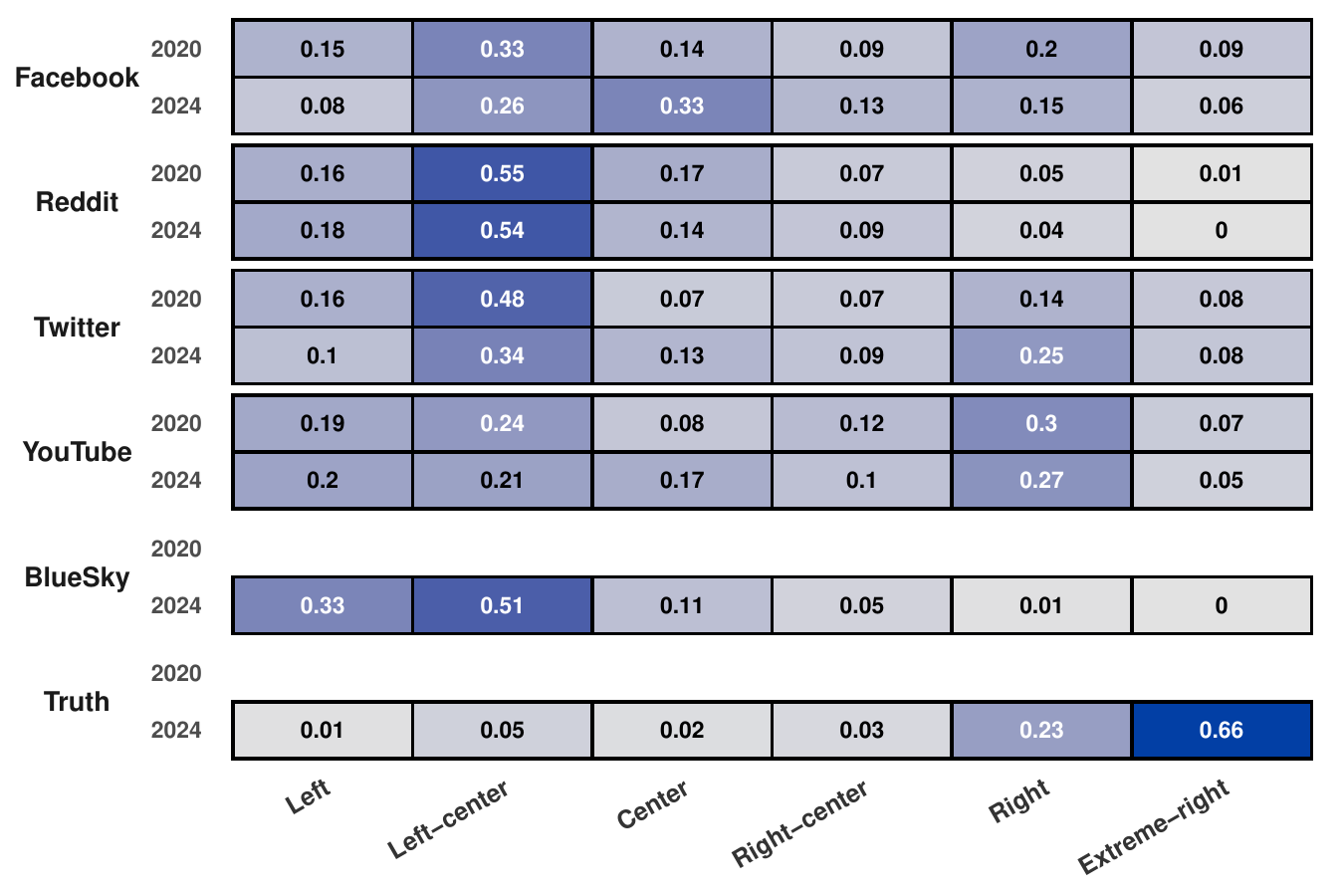}
    \caption{\textbf{Fraction of shared URLs linking to external news domains across platforms, stratified by political leaning, on the temporally unified data sets.}}
    \label{fig:shares_common_window}
\end{figure}

\begin{figure}[h]
    \centering
    \includegraphics[width=\columnwidth]{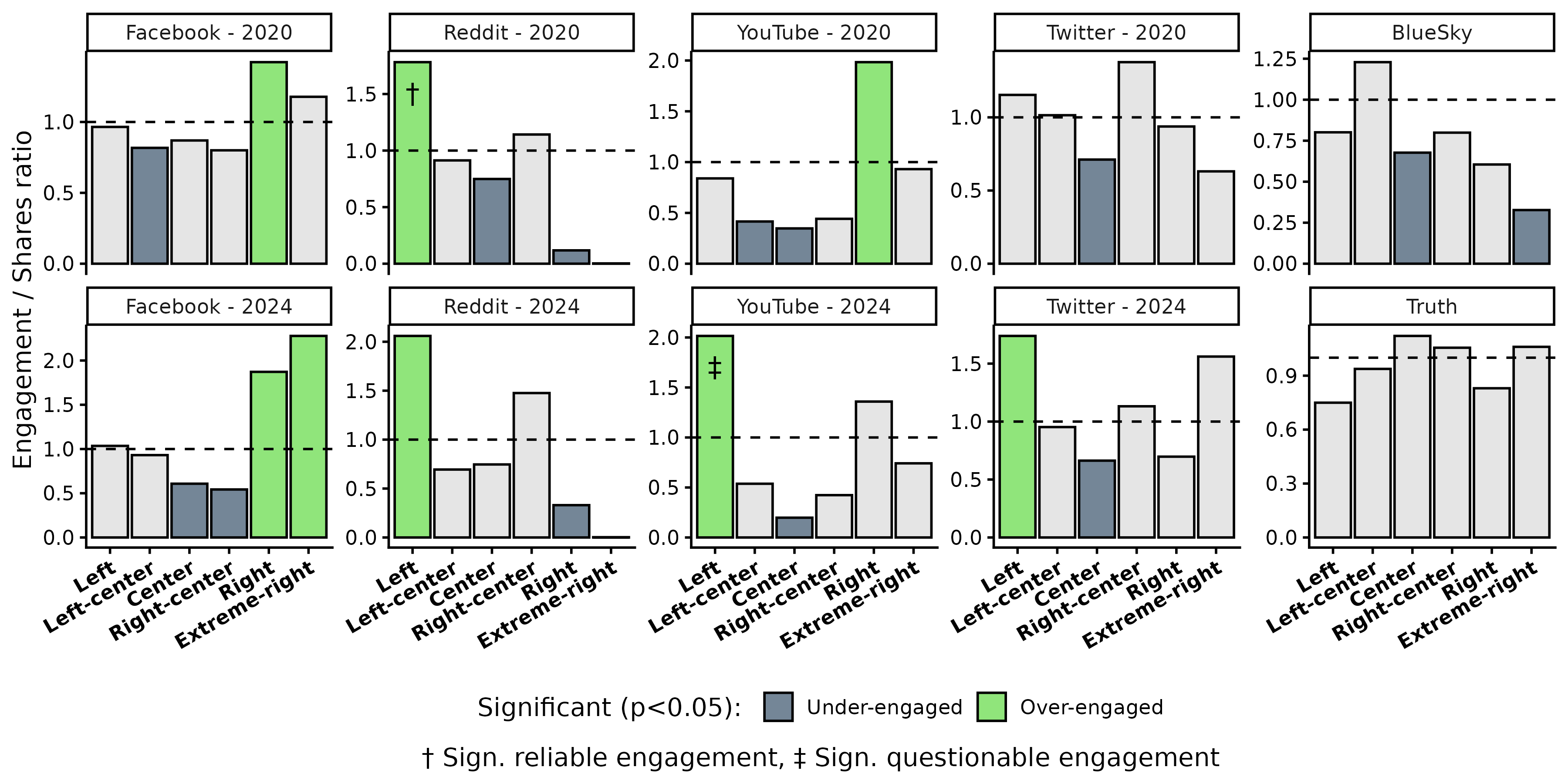}
    \caption{\textbf{Engagement-to-share ratio $R_{p,\ell}$ across temporally unified data sets.}}
    \label{fig:random_eng_common_window}
\end{figure}

A key tradeoff in the analysis concerns preserving the full temporal extent of the data versus ensuring consistent coverage across platforms. In the main analysis, we retain the complete datasets to avoid discarding potentially informative data (Table~\ref{tab:delta_edgelists}). To assess the impact of this choice, we replicate the analysis using a common overlapping time window for each year (Figure~\ref{fig:common_windows}), enabling a more directly comparable setting across platforms. Reducing the platforms to the common window bears little to no impact on the distribution of sources shared (Figure \ref{fig:shares_common_window}), and the allocation of engagement (Figure \ref{fig:random_eng_common_window}), confirming that the main results are robust to differences in temporal coverage.

\begin{table}[!h]
\caption{Difference in values between the original and unified timeframe datasets.
Columns report the number of labeled URL links in the original datasets ($n_u$) and in the datasets restricted to a unified timeframe ($n_u^{(U)}$), together with their difference ($\Delta n_u$). Analogously, we report the number of unique labeled domains in the original datasets ($n_d$) and in the unified timeframe datasets ($n_d^{(U)}$), together with their difference ($\Delta n_d$).}
\centering
\begin{tabular}{lcccccc}
\toprule
Platform & $n_u$ & $n_u^{(U)}$ &  $\Delta n_u$ & $n_d$ & $n_d^{(U)}$ & $\Delta n_d$\\
\midrule
Facebook (2020) & 5,167,018 & 4,416,384 & -750,634 & 4,084 & 3,980 & -104\\
Facebook (2024) & 662,520 & 514,637 & -147,883 & 3,018 & 2,916 & -102\\
\addlinespace
Reddit (2020) & 372,196 & 172,513 & -199,683 & 2,571 & 2,122 & -449\\
Reddit (2024) & 112,844 & 56,273 & -56,571 & 1,777 & 1,347 & -430\\
\addlinespace
Twitter (2020) & 69,579,321 & 69,579,321 & 0 & 3,555 & 3,555 & 0\\
Twitter (2024) & 1,450,124 & 1,329,488 & -120,636 & 3,642 & 3,602 & -40\\
\addlinespace
YouTube (2020) & 117,419 & 87,381 & -30,038 & 1,267 & 1,076 & -191\\
YouTube (2024) & 583,011 & 409,433 & -173,578 & 1,416 & 1,256 & -160\\
\addlinespace
Bluesky & 305,448 & 82,799 & -222,649 & 2,227 & 1,425 & -802\\
Truth & 171,748 & 145,500 & -26,248 & 1,521 & 1,322 & -199\\
\bottomrule
\end{tabular}
\label{tab:delta_edgelists}
\end{table}

\end{document}